\documentclass[aps,prb,superscriptaddress,amsmath,amssymb,preprint]{revtex4-2}

\usepackage{graphicx}
\usepackage{dcolumn}
\usepackage{bm}
\usepackage{stmaryrd}
\usepackage{upgreek}
\makeatletter

\begin{document}

\preprint{}

\title[A compact and tunable forward coupler based on high-impedance superconducting nanowires]{A compact and tunable forward coupler based on high-impedance superconducting nanowires}

\author{Marco Colangelo}
\thanks{These two authors contributed equally}
\affiliation{Research Laboratory of Electronics, Massachusetts Institute of Technology, Cambridge, Massachusetts 02139, USA}

\author{Di Zhu}
\thanks{These two authors contributed equally}
\affiliation{Research Laboratory of Electronics, Massachusetts Institute of Technology, Cambridge, Massachusetts 02139, USA}
 \affiliation{John A. Paulson School of Engineering and Applied Science, Harvard University, Cambridge, Massachusetts 02138, USA}

\author{Daniel F. Santavicca}
\affiliation{Department of Physics, University of North Florida, Jacksonville, Florida 32224, USA}
\author{Brenden A. Butters}
\affiliation{Research Laboratory of Electronics, Massachusetts Institute of Technology, Cambridge, Massachusetts 02139, USA}
\author{Joshua C. Bienfang}
\affiliation{National Institute of Standards and Technology, Gaithersburg, Maryland 20899, USA}

\author{Karl K. Berggren}
\email{berggren@mit.edu}
\affiliation{Research Laboratory of Electronics, Massachusetts Institute of Technology, Cambridge, Massachusetts 02139, USA}

\date{\today}
\begin{abstract}
Developing compact, low-dissipation, cryogenic-compatible microwave electronics is essential for scaling up low-temperature quantum computing systems. In this paper, we demonstrate an ultra-compact microwave directional forward coupler based on high-impedance slow-wave superconducting-nanowire transmission lines. The coupling section of the fabricated device has a footprint of $416\,\mathrm{\upmu m^2}$. At 4.753 GHz,  the input signal couples equally to the through port and forward-coupling port (50:50) at $-6.7\,\mathrm{dB}$ with $-13.5\,\mathrm{dB}$ isolation. The coupling ratio can be controlled with DC bias current or temperature by exploiting the dependence of the kinetic inductance on these quantities. The material and fabrication-process are suitable for direct integration with superconducting circuits, providing a practical solution to the signal distribution bottlenecks in developing large-scale quantum computers.

\end{abstract}


\maketitle

The scalability of superconducting quantum systems is constrained by the distribution of microwave signals to the quantum processors \citep{blais2020quantum}. Within current designs, each qubit is individually wired for readout and control \citep{arute2019quantum,blais2020quantum}, entailing an increasing number of devices and cables as the size of the circuit is increased. Inevitably, the present approach will lead to challenges in packaging, routing, thermalization, and footprint \citep{gambetta2017building}. To realize large-scale circuits with thousands of qubits, most of the microwave electronics will need to be integrated on-chip \citep{blais2020quantum,gambetta2017building}, necessitating the development of miniaturized low-power, low-dissipation RF devices. More broadly, a small-footprint cryogenic microwave electronics platform is also required for the advancement of several other applications relying on processing electrical signals at low temperature, such as single-photon detection \citep{holzman2019superconducting}, superconducting quantum interference device (SQUID) magnetometry \citep{fagaly2006superconducting,buchner2018tutorial}, and radio astronomy \citep{zmuidzinas2012superconducting,mchugh2012readout} .

\begin{figure*}
    \centering
    \includegraphics{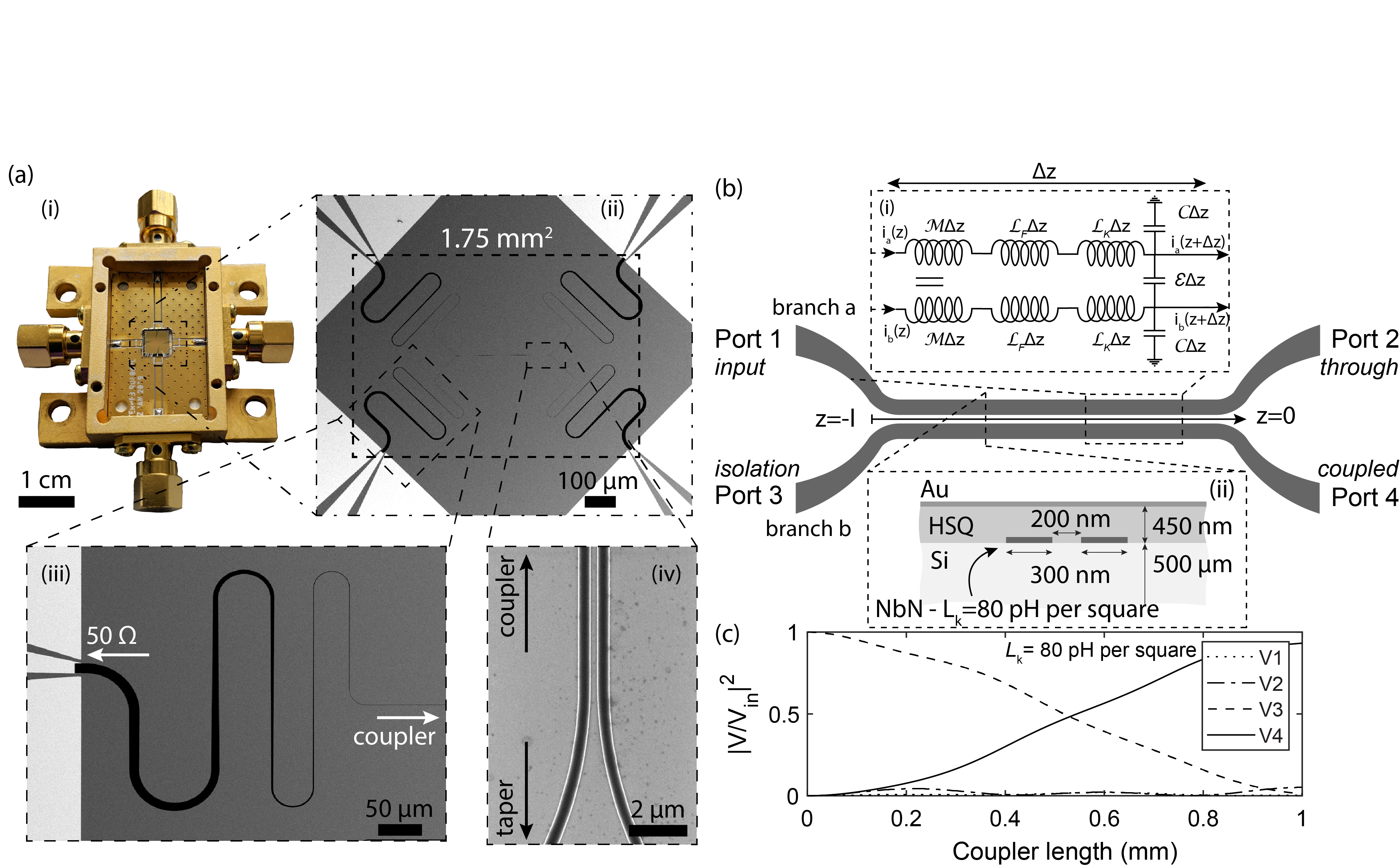}
    \caption{Superconducting-nanowire directional forward coupler: device and model. (a) Optical and scanning electron micrographs (SEM) of the fabricated device: (i) device die mounted in the radio-frequency (RF) testing box; (ii) SEM showing the full four-port device before the fabrication of the dielectric spacer and the top ground; (iii) close-up SEM of the impedance-matching taper; (iv) close-up SEM of the coupled nanowire transmission lines. (b) Geometry, model, and implementation of coupler: two nanowire transmission lines brought together for a coupling length $l$: (i) analytical model: coupled LC ladder, where we explicitly separated the geometric (Faraday) component of the total line inductance per unit length (p.u.l.), $\mathcal{L}_{\mathrm{F}}$, from the kinetic inductance p.u.l. $\mathcal{L}_{\mathrm{k}}$; $\mathcal{C}$ is the capacitance p.u.l. of each line; $\mathcal{E}$ is the coupling capacitance p.u.l.; $\mathcal{M}$ is the mutual inductance p.u.l.; (ii) Implementation as side-coupled striplines. In the analytical model, the gold layers (Au) have been treated as perfect electric conductors (PEC) (c) Simulation of the port voltages versus coupler length at $5\,\mathrm{GHz}$ for 50:50 forward coupling ($L_{\mathrm{k}}=80\,\mathrm{pH}$ per square).}
    \label{fig2}
\end{figure*}

Many of the proposals to address scalability in superconducting circuits face several challenges in satisfying the requirements for on-chip integrability. Devices based on semiconductors \citep{bardin201929,patra2017cryo,hornibrook2015cryogenic,borodulin2019operation,ruffino20196} either dissipate too much power to be operated at a few milliKelvin \citep{blais2020quantum} or are made from unconventional materials for which integration with superconductors has not yet been demonstrated. Among the superconducting solutions, $50\,\Omega$ transmission-line-based devices \citep{pechal2016superconducting,ku2010design} require too large a footprint for large-scale integration. Josephson junction (JJ) electronics are a natural candidate for integration with JJ-based quantum processors \citep{leonard2019digital,brummer2010phase,chapman2016general,naaman2017josephson,naaman2016chip,abdo2017gyrator}, but they can be challenging to manufacture and require magnetic shielding.

Recently, superconducting nanowires have emerged as an alternative approach to realize ultra-compact microwave devices \citep{lowell2016thin,wagner2019demonstration}. The high kinetic inductivity achieved with disordered superconducting thin-films provides an effective means of realizing extremely high characteristic-impedance transmission lines with zero DC resistance, minimal microwave dissipation, slow phase velocity (high effective refractive index), and very small footprint. The native high-impedance, high-index operation generates electromagnetically-protected microwave environments that could well interface to superinductor-based qubits \citep{masluk2012microwave,groszkowski2018coherence,kjaergaard2019superconducting,grunhaupt2019granular}. The compatibility with these applications is guaranteed by the rather conventional materials and by the few-layer fabrication process \citep{hazard2019nanowire,niepce2019high}, making the devices realized with superconducting nanowires a viable solution to the signal distribution bottlenecks of quantum computers.

Conventional directional coupler modules split, combine and distribute microwave fields to the subsequent processing layers or to the readout \citep{ku2010design,gu2017microwave,krinner2019engineering}. They carry out essential processing tasks but take up a significant volume inside the cryostat. Here, we use high kinetic-inductance superconductors to demonstrate a compact high-impedance directional coupler.  Our miniaturized coupler is based on niobium nitride (NbN) superconducting nanowire side-coupled striplines embedded in a multilayer dieletric stack. Fig. \ref{fig2}(a) shows optical and scanning electron micrographs (SEM) of the fabricated device. We achieve forward coupling in the GHz range in an extremely reduced footprint and with impedance matching flexibility. We further demonstrate that the non-linear dependence of the nanowire's kinetic inductance on DC current and temperature allow the microwave properties of the coupler to be tuned. We suggest this device may find immediate application in superconducting quantum computing systems and in other low-temperature applications where small-footprint on-chip coupling at microwave frequency is needed.

Our device is designed following a traditional coupled-line architecture (Fig. \ref{fig2}(b)) where two superconducting nanowires are brought together for a coupling length $l$. The structure can be modeled as a coupled LC ladder (Fig. \ref{fig2}(b)(i)) using a standard coupled-mode formalism \citep{garg2013microstrip,pozar2005microwave,tripathi1975asymmetric}, with some modifications to capture the kinetic-inductive transmission line \citep{zhu2019microwave}. To reflect the high-inductivity behavior of nanowires, we explicitly separate the kinetic contribution ($\mathcal{L}_\mathrm{k}$), from the geometric (Faraday) contribution ($\mathcal{L}_\mathrm{F}$) to the total line inductance per unit length. The coupling produces mode splitting into common ($c$) and differential ($\pi$) modes, with different effective indices and propagation constants ($\beta_\mathrm{c}$ and $\beta_\mathrm{\pi}$). In a transmission line, a sinusoidal signal with frequency $\omega$ is a superposition of these eigenmodes and energy is continuously transferred between the two lines with a periodicity $l_{\mathrm{\pi}}=\pi / \Delta \beta = \pi / \left( \beta_{\pi} - \beta_{c} \right)$. A section of coupled transmission line with a length that is an odd-integer multiple of $l_{\mathrm{\pi}}/2$ can perform, in the ideal case, $3\,\mathrm{dB}$ forward coupling at $\omega$. In RF transmission lines made of conventional materials, the splitting in propagation constant $\Delta \beta$ is relatively small and the minimum length required to achieve forward coupling (at a target frequency) is relatively large. Therefore, it is generally more convenient to exploit the difference in the characteristic impedance of the eigenmodes to realize low-coupling backward directional couplers \citep{pozar2005microwave,morgan2003octave}. With coupled superconducting nanowires, the combination of high-kinetic inductance lines, high coupling capacitance, and low loss boosts $\Delta \beta \propto \sqrt{\mathcal{L}_\mathrm{k}\mathcal{C}}\sqrt{1+2\mathcal{E/C}}$ and allows forward coupling in a small footprint ($l_\pi$ is relatively small). Moreover, backward reflections can be minimized by gradually tapering the exits from the coupled-line section (Fig. \ref{fig2}(a)(iv)). See Supplemental Material (SM) for the full derivation of the model.

To practically illustrate this concept, we consider the side-coupled stripline implementation, shown in Fig. \ref{fig2}(b)(ii), which we used to realize our device. The lines are made of a $300\,\mathrm{nm}$-wide, $7\,\mathrm{nm}$-thick NbN nanowire with a sheet kinetic inductivity of $L_{\mathrm{k}}=80\,\mathrm{pH}$ per square, separated by a $200\,\mathrm{nm}$ wide gap. The structures are patterned on silicon and referenced to a $60\,\mathrm{nm}$ topside gold ground plane through a $450\,\mathrm{nm}$ thick hydrogen silsequioxane (HSQ) dielectric layer, with $\varepsilon_\mathrm{r}=2.9$ \citep{Maier2001,zhu2018scalable}. In this microwave environment, the simulated characteristic impedance is $Z_0=1446\,\mathrm{\Omega}$ and the effective index $n_\mathrm{eff}=54.5$, which reduces the phase velocity to $1.8\%$ of $c$ and highly compresses the guided wavelength. See SM for details on the simulations. For clarity, the port naming convention is shown in Fig. \ref{fig2}(b). Fig. \ref{fig2}(c) shows that a $5\,\mathrm{GHz}$ signal injected through the input port of the coupling section takes $l_{\pi/2}=\pi/\left(2 \Delta \beta \right)=539\,\mathrm{\upmu m}$ to forward couple 50\% of the power to the other branch. Compared to the same structure realized with conventional conductors \citep{ikalainen1987wide}, this coupling section achieves almost two orders of magnitude footprint reduction (see SM).

\begin{figure*}
    \centering
    \includegraphics{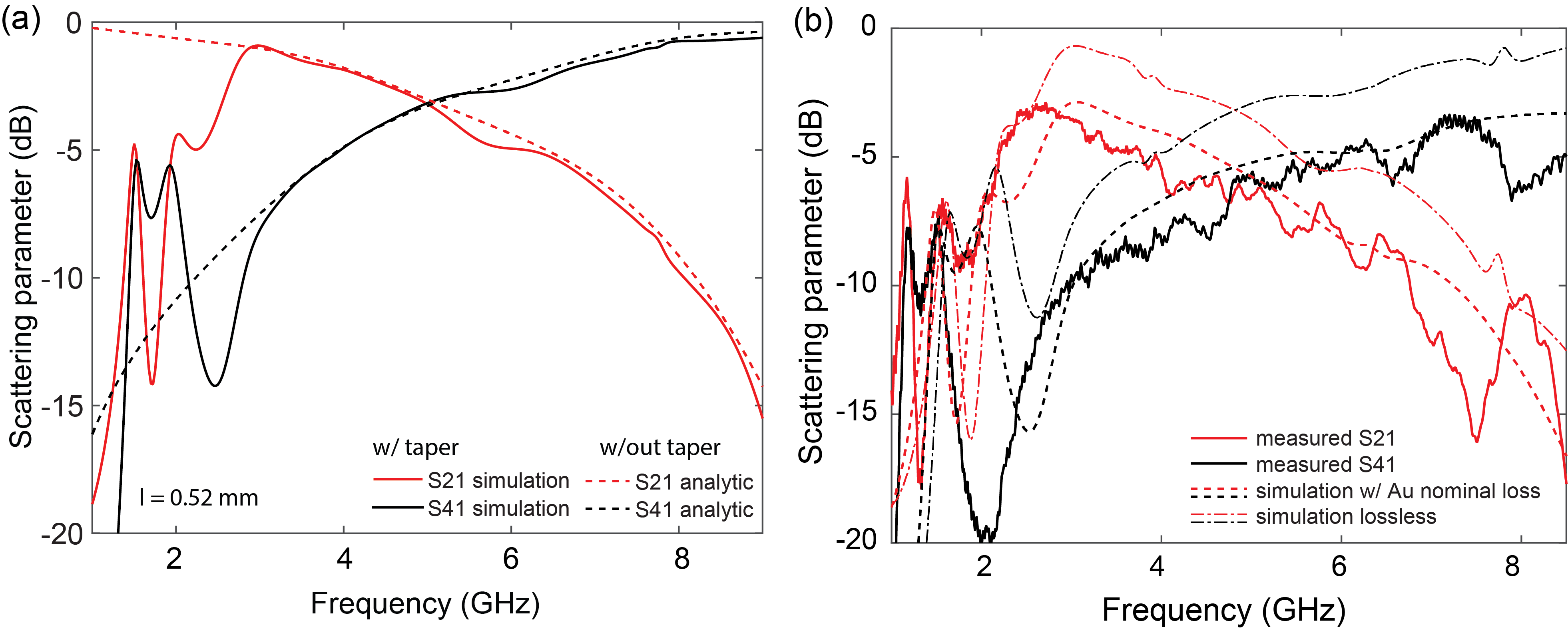}
    \caption{Microwave response of the microwave directional forward coupler. (a) Analytical modeling of the coupling section (without taper) and simulation of the impedance-matched coupler. Above $3\,\mathrm{GHz}$ the curves match, showing that the impedance-matching taper, interfacing the high-impedance coupling section to the $50\,\mathrm{\Omega}$ RF electronics, does not disturb the coupling operation. (b) Measured microwave response of the fabricated device at $1.3\,\mathrm{K}$. We compare the data to simulation, corrected for fabrication non-idealities, and including normal conductor losses from the Au ground plane and CPW feedlines. The balanced forward coupling is observed at $4.753\,\mathrm{GHz}$.}
    \label{fig2p1}
\end{figure*}

The device was fabricated with parameters based on the modeling results. See SM for details on the fabrication. Fig. \ref{fig2}(a) shows micrographs of the device before the fabrication of the dielectric spacer and the top ground. In the final design we used a coupling length $l=520\,\mathrm{\upmu m}$ and we included four $2.5\,\mathrm{GHz}$ high-pass Klopfenstein impedance-matching tapers (Fig. \ref{fig2}(a)(iii)) \citep{klopfenstein1956transmission,zhu2019superconducting,zhu2018scalable,zhao2017single,zhu2019resolving}.  Coplanar waveguide (CPW) signal feed lines ($60\,\mathrm{nm}$ gold) were also fabricated to allow wire-bonding and packaging. The footprint of the full superconducting structure (as outlined in Fig.  \ref{fig2}(a)(ii)) was $1.75\,\mathrm{mm^2}$, while the high-impedance coupling section only occupied $416\,\mathrm{\upmu m}^2$.

The addition of the tapers, while increasing the total footprint of the device, improves  flexibility, affording the possibility to interface the high-impedance coupling section to lower-impedance environments without disturbing the forward coupling operation. In this case we interfaced the high-impedance nanowire transmission line coupler to $50\,\mathrm{\Omega}$ RF electronics. Fig.  \ref{fig2p1}(a) shows that, in the taper passband ($f \geq 3\,\mathrm{GHz}$), the simulation results (full wave) of the impedance-matched coupler correctly reproduce the analytic calculation of the high-impedance coupling section response. The 50\% coupling point is at $4.99\,\mathrm{GHz}$, with the isolation parameter at $-21.9\,\mathrm{dB}$. See SM for details on the simulations.

We measured the microwave response of the fabricated coupler at $1.3\,\mathrm{K}$ with a vector network analyzer providing effective signal power lower than $-60\,\mathrm{dBm}$. In the same cooldown, we calibrated the cable and connector loss (cryostat to device box inputs/outputs) to scale the measured data. Fig.  \ref{fig2p1}(b) shows that the forward coupling behavior is observed. At $4.753\,\mathrm{GHz}$, the input signal from Port 1 couples equally to the through port (Port 2) and forward-coupling port (Port 4) with a level of $-6.7\,\mathrm{dB}$, and the isolation parameter $S31=-13.5\,\mathrm{dB}$. For comparison, we show simulations of the impedance-matched coupler, including the response of the feed lines, and corrected to account for fabrication non-idealities (see SM for additional details). The slight discrepancy ($\approx 500\,\mathrm{MHz}$) between the measured and simulated coupling frequency can be attributed to the uncertainties in the device parameters. For example, the fabrication process, consisting of several lithographic and etching steps, may induce a degradation of the film leading to an increase in the kinetic inductance that would explain this observation. We attribute the inconsistency in the magnitude of the S parameters to backward reflections, device-level conductor losses, and to additional contributions that were not accounted for in the calibration. In this experiment, the calibration does not account for losses and reflection from the sample holder PCB and wire bonds. The agreement with the simulation improves when we include normal conductor losses for the Au layers. More details can be found in the SM. The isolation parameter is at a significantly different level from the expected value. This discrepancy might be caused by factors such as the impedance-matching taper deviating from the prescribed design, with sub-optimal impedance matching and additional backward reflections, and the specific full device simulation not including element-to-element transitions (e.g. stripline to CPW). See SM for additional details.

\begin{figure*}
    \centering
    \includegraphics{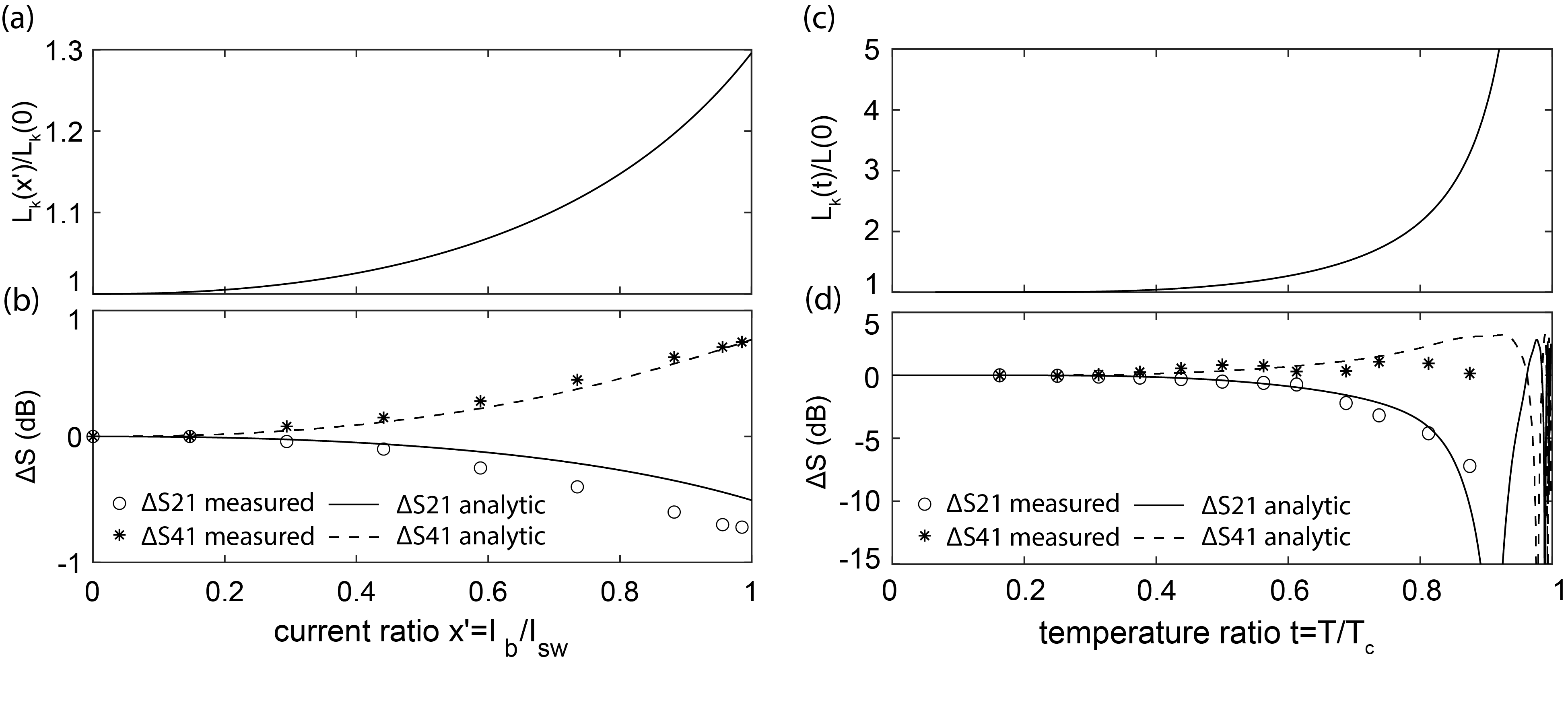}
    \caption{Non-linear dependence of the kinetic inductance $L_\mathrm{k}(x'=I_\mathrm{b}/I_\mathrm{sw},t=T/T_\mathrm{c})$ and coupler response tunability at $=4.75\,\mathrm{GHz}$ with bias current and temperature. (a) Kinetic inductance's dependence on current, for $I_{\mathrm{sw}}/I_{\mathrm{d}}=0.7$, according to the fast relaxation model with $\alpha=2.27$. (b) Tunability of the transmission (S21) and coupling (S41) response as a function of the current (normalized by $I_{\mathrm{sw}}$) supplied through the isolation port. The measured data (symbols) well match our analytical model (lines) based on coupled-mode formulation with current-dependent kinetic inductance. (c) Kinetic inductance's dependence on temperature calculated from the explicit solution of the temperature dependence of the superconducting gap $\Delta(T)$, with $N(0)V=0.32$ and $T_{\mathrm{c}}=8\,\mathrm{K}$ (see SM). (d) Modulation of transmission and coupling parameters, as a function of the cryostat temperature (normalized by $T_{\mathrm{c}}$). The measured data (symbols) match our analytical model (lines) based on the coupled-mode formulation with temperature-dependent kinetic inductance.}
    \label{fig5}
\end{figure*}
The microwave response was characterized with effective injected currents much smaller than the device depairing current $I_\mathrm{d}$ (see SM). In this small-signal condition the device response is independent of the applied microwave power. However, the kinetic inductance strongly depends on carrier density, which can be tuned with current ($I_\mathrm{b}$) and with temperature ($T$): $L_k=L_k(x'=I_\mathrm{b}/I_\mathrm{sw},t=T/T_\mathrm{c})$, where $I_{\mathrm{sw}}$ is the switching current and $T_\mathrm{c}$ is the critical temperature. Leveraging these dependencies affords the possibility of dynamically modulating the microwave characteristics of the modes in the lines, creating active, tunable devices. 
When the bias current in the nanowire approaches the depairing current ($x \rightarrow 1$), the kinetic inductance diverges hyperbolically $L_\mathrm{k}(x) \propto (1-x^{\alpha})^{-1/\alpha}$, with $\alpha$ determined by the operating temperature of the device \citep{clem2012kinetic}. Due to fabrication imperfections, superconducting nanowire devices similar to the ones described in this paper typically can only reach a fraction of the depairing current, with the switching current $I_{sw} \approx 70\%\,I_{d}$ \citep{frasca2019determining}. Fig. \ref{fig5}(a) shows that with this assumption and given the measurement condition ($\alpha = 2.27$) \citep{clem2012kinetic}, a $30\%$ theoretical maximum increment of the kinetic inductance might be expected. In the coupled-line architecture the two nanowires are galvanically isolated. Hence, the kinetic inductance of each nanowire can be tuned independently.  We characterized the device tunability at $4.75\,\mathrm{GHz}$ by biasing the coupler through the isolation port and measuring the coupled and transmitted powers. Fig. \ref{fig5}(b) shows the change of scattering parameters $\Delta S$ with the applied current, in agreement with analytical modeling. The increase of kinetic inductance increases S41 and reduces S21 at the original equal-coupling frequency, which in turn shifts the 50:50 coupling point to a lower frequency (see SM).

When changing the operating temperature, a much wider variation of the kinetic inductance, and hence of the scattering parameters, can be achieved. Fig. \ref{fig5}(c) shows the expected theoretical temperature dependence of the kinetic inductance obtained through numerical calculation of the superconducting gap, and the corresponding kinetic inductance  $L_\mathrm{k}(T) \propto  \left[\Delta(T)\tanh \left(\Delta(T)/T\right)\right]^{-1}$ \citep{santavicca2016microwave,tinkham2004introduction}, using characteristic values for NbN \citep{polakovic2018room}. We characterized the coupling tunability by varying the base temperature of the cryostat and measuring the S21 and S41, at $4.75\,\mathrm{GHz}$. With this measurement setup, the variation of the temperature modulates the kinetic inductance of both the nanowires. Fig.  \ref{fig5}(d) shows tunability of the parameters with temperature and is in fair agreement with the model. The observed discrepancies with the theoretical curves are attributed to the uncertainty in the modeling parameters used for the calculation of $\Delta(T)$. Moreover, as the calculation only includes the coupling section, the impact of other elements of the device (e.g. tapers) was not captured. Note that when the kinetic inductance is far from the design value, the microwave characteristics of the taper can significantly diverge from the intended behavior.

The device presented in this work achieves 50:50 forward coupling in a dramatically reduced footprint by exploiting the properties of high-inductance superconducting nanowire transmission lines. The coupling section, as configured here, occupies only $416\,\mathrm{\upmu m}^2$ and can be integrated as-is in high-impedance environment circuits with $Z_0\approx 1.5\,\mathrm{k\Omega}$. Moreover, as the characteristic-impedance depends strongly on the device geometry, the device can be matched to a wide variety of high-impedance environments by redesigning the coupling section's width and length, while keeping a small footprint. As mentioned above, the possibility of performing impedance matching using tapered structures allows one to interface the directional coupler to lower-impedance environment as well. The total occupied footprint, even including the tapers, is still lower than other conventional normal-conductor coupler designs, such as hybrid, Lange, or parallel lines \citep{pozar2005microwave}, which require $\approx 10\,\mathrm{mm^2}$. Additional footprint reduction may be achieved by optimizing the packing of current layouts, or by using a higher effective-index transmission-line architecture \citep{zhu2019superconducting}. Alternatively, a broadside-coupler architecture might be realized allowing an increase of the capacitive coupling and a further reduction of the coupling length (SM).  

The model developed to support the design of the device is in agreement with the measured response. The disagreement with the measured data, observed in the magnitude of the scattering parameters, are partially due to backward reflections and to device-level losses which are not accounted for in the system calibration. The effects induced by the use of normal conductors (feed lines and ground plane) or lossy dielectrics can be addressed by redesigning the material stack and adapting the layout. The device model also does not include the power-dependent non-linear effects that might play an additional role when driving the coupler in the highly non-linear kinetic-inductance regime. This could contribute to the discrepancies observed when testing the coupler for $t\rightarrow 1$, where t is the reduced temperature. The study of the non-linearities in high-kinetic-inductance transmission lines is beyond the scope of this paper. 

The tunability of the coupling parameters opens up the opportunity for the realization of high-impedance tunable microwave devices. For example, a high-impedance single-pole double-throw non-linear switch could be realized based on this coupler architecture. Similarly, tunability with temperature may become practical if a heater element is fabricated in close proximity to the coupling section.

We suggest this device may find application in existing superconducting quantum architectures, where the integration of superconducting nanowires, in the form of superinductors, has already been demonstrated \citep{hazard2019nanowire,niepce2019high,ku2010superconducting}. A high-impedance cryogenic tunable coupler could be used for tunable qubit-qubit coupling \citep{krantz2019quantum}, on-chip integration of novel readout techniques \citep{abdo2019active}, and on-chip signal processing and multiplexing, drastically reducing the necessary wiring from couplers and splitter on the higher temperature stage to the processor at mK. In the effort to scale the size of single-photon detectors arrays, this device could be used to implement architectures based on frequency-multiplexing readout \citep{doerner2017frequency,sinclair2019demonstration}. Further development of this nanowire-based technology may lead to the realization of a family of ultra-compact microwave devices that form the basis of a new superconducting nanowire monolithic microwave integrated circuit technology.

See Supplemental Material for more details on modeling, simulation, fabrication and measurements. All files related to a published paper are stored as a single deposit and assigned a Supplemental Material URL. This URL appears in the article’s reference list.

\begin{acknowledgments}
We thank K. O'Brien, K. Peng, M. Naghiloo for helpful discussions. We thank E. Toomey, J. Simonaitis, and M. Bionta for critical reading of the manuscript. Support for this work was provided in part by the National Science Foundation grants under contract No. ECCS-2000743 and ECCS-2000778 (UNF), and by the Army Research Office (ARO) under Cooperative Agreement Number W911NF-16-2-0192. The views and conclusions contained in this document are those of the authors and should not be interpreted as representing the official policies, either expressed or implied, of the Army Research Office or the U.S. Government. The U.S. Government is authorized to reproduce and distribute reprints for Government purposes notwithstanding any copyright notation herein. Di Zhu was supported by the National Science Scholarship from A*STAR, Singapore, and Harvard Quantum Initiative Postdoctoral Fellowship.

\end{acknowledgments}

\bibliography{BiblioXY}

\end{document}


\title[A compact and tunable forward coupler based on high-impedance superconducting nanowires]{Supplemental Materials: A compact and tunable forward coupler based on high-impedance superconducting nanowires}

\author{Marco Colangelo}
\thanks{These two authors contributed equally}
\affiliation{Research Laboratory of Electronics, Massachusetts Institute of Technology, Cambridge, Massachusetts 02139, USA}

\author{Di Zhu}
\thanks{These two authors contributed equally}
\affiliation{Research Laboratory of Electronics, Massachusetts Institute of Technology, Cambridge, Massachusetts 02139, USA}
 \affiliation{John A. Paulson School of Engineering and Applied Science, Harvard University, Cambridge, Massachusetts 02138, USA}

\author{Daniel F. Santavicca}
\affiliation{Department of Physics, University of North Florida, Jacksonville, Florida 32224, USA}
\author{Brenden A. Butters}
\affiliation{Research Laboratory of Electronics, Massachusetts Institute of Technology, Cambridge, Massachusetts 02139, USA}
\author{Joshua C. Bienfang}
\affiliation{National Institute of Standards and Technology, Gaithersburg, Maryland 20899, USA}

\author{Karl K. Berggren}
\email{berggren@mit.edu}
\affiliation{Research Laboratory of Electronics, Massachusetts Institute of Technology, Cambridge, Massachusetts 02139, USA}

\keywords{Suggested keywords}
\maketitle

\tableofcontents
\newpage

\section{Analytic model}
In this section we describe the analytical modeling of the nanowire coupler, originally presented in \citep{zhu2019microwave}. It is based on a coupled-mode formalism \citep{garg2013microstrip,tripathi1975asymmetric} adapted to explicitly include the kinetic contribution to the total line inductance.

We consider here the schematic shown in the main text in Fig. 1(b). The inductance per unit length (\textit{p.u.l.}) of each line has been separated into two components $\mathcal{L}_{a,b}=\mathcal{L}_{\mathrm{Fa,Fb}} + \mathcal{L}_\mathrm{ka,kb}$, which are the geometric (Faraday) and kinetic inductance contributions, respectively.  $\mathcal{M}$ and $\mathcal{E}$ are the \textit{p.u.l.} mutual inductance and coupling capacitance between the two lines. $\mathcal{C}_{a,b}$ are the \textit{p.u.l.} self-capacitances, corrected for the fringing field component \citep{bedair1984characteristics}. We can write the coupled telegrapher's equations:
\begin{subequations}
    \begin{align}
    -\partial_z \begin{bmatrix}i_a\\i_b\end{bmatrix} &= \begin{bmatrix}\mathcal{C}_a +\mathcal{E}& -\mathcal{E}\\ -\mathcal{E} & \mathcal{C}_b + \mathcal{E}\end{bmatrix}\partial_t\begin{bmatrix}v_a\\v_b\end{bmatrix}\label{eq:coupledTelegrapherA}\\    
    -\partial_z \begin{bmatrix}v_a\\v_b\end{bmatrix} &= \begin{bmatrix}\mathcal{L}_a +\mathcal{M}& -\mathcal{M}\\ -\mathcal{M} & \mathcal{L}_b + \mathcal{M}\end{bmatrix}\partial_t\begin{bmatrix}i_a\\i_b\end{bmatrix}\label{eq:coupledTelegrapherB}.
    \end{align}
\end{subequations}
Taking the partial derivative with respect to $z$ of Eq. (\ref{eq:coupledTelegrapherB}) and substituting in Eq. (\ref{eq:coupledTelegrapherA}) we have:
\begin{align}
\begin{split}
\partial^2_z \begin{bmatrix}v_a\\v_b\end{bmatrix} =\begin{bmatrix}\alpha_a&\gamma_a\\\gamma_b&\alpha_b\end{bmatrix}\partial^2_t\begin{bmatrix}v_a\\v_b\end{bmatrix}\label{eq:wave_eq}
\end{split}
\end{align}
with
\begin{align}
\alpha_a&=(\mathcal{L}_a+\mathcal{M})(\mathcal{C}_a+\mathcal{E})+\mathcal{M}\mathcal{E}\\
\gamma_a &= -\mathcal{E}(\mathcal{L}_a+\mathcal{M})-\mathcal{M}(\mathcal{C}_b+\mathcal{E})\\
\alpha_b &= (\mathcal{L}_b+\mathcal{M})(\mathcal{C}_b+\mathcal{E})+\mathcal{E}\mathcal{M} \\
\gamma_b &=-\mathcal{M}(\mathcal{C}_a+\mathcal{E})-\mathcal{E}(\mathcal{L}_b+\mathcal{M}).
\end{align}
When two transmission lines are brought in close proximity, their coupling produces mode splitting into common ($c$) and differential ($\pi$) modes, having different effective indices and propagation constants. Assuming the voltages in the two lines $v_{a,b}(z,t)$ propagate in the form of $v_{a,b} = V_{a,b} e^{j\omega t - j\beta z}$ for the eigenmodes, we can solve the dispersion relation

\begin{equation}
\frac{\beta_{c,\pi}^2}{\omega^2} = \frac{(\alpha_a+\alpha_b)\pm\sqrt{(\alpha_a-\alpha_b)^2+4\gamma_a\gamma_b}}{2},\label{eq:eigen_mode}
\end{equation}
and for the two eigenmodes, the voltage ratios on the two lines are
\begin{equation}
R_{c,\pi} = \frac{v_b}{v_a} =\frac{\alpha_b-\alpha_a\pm\sqrt{(\alpha_a-\alpha_b)^2+4\gamma_a\gamma_b}}{2\gamma_a}.
\end{equation}
We now derive the general solution for the voltages on the lines in terms of forward and backward propagating waves for the $c$ and $\pi$ modes:
\begin{align}
\begin{split}\label{Va}
V_a(z)=&A_1 e^ {-j\beta_c z} + A_2 e^ {j\beta_c z} + \\ &A_3 e^ {-j\beta_\pi z} + A_4 e^ {j\beta_\pi z}
\end{split}
\\
\begin{split}\label{Vb}
V_b(z)=&A_1 R_c e^ {-j\beta_c z} + A_2 R_c e^ {j\beta_c z} + \\&A_3 R_\pi e^ {-j\beta_\pi z} + A_4 R_\pi e^ {j\beta_\pi z}.
\end{split}
\end{align}

The currents on the line can be obtained by substituting Eq.\ref{Va} and Eq.\ref{Vb} in Eq.\ref{eq:coupledTelegrapherB}:
\begin{align}
\begin{split}
I_a(z)=&\frac{A_1}{Z_{c,a}} e^{-j\beta_c z} - \frac{A_2}{Z_{c,a}} e^{j\beta_c z} +\\& \frac{A_3}{Z_{\pi,a}} e^ {-j\beta_\pi z}- \frac{A_4}{Z_{\pi,a}} e^ {j\beta_\pi z}
\end{split}
\\
\begin{split}
I_b(z)=&\frac{R_c A_1}{ Z_{c,b}} e^{-j\beta_c z} - \frac{R_c A_2}{Z_{c,b}} e^{j\beta_c z} + \\&\frac{R_\pi A_3}{ Z_{\pi,b}}e^ {-j\beta_\pi z}- \frac{ R_\pi A_4}{ Z_{\pi,b}} e^ {j\beta_\pi z}\end{split}
\end{align}
where $Z_{c,a,b}$ and $Z_{\pi,a,b}$ denotes the common and differential mode impedances \citep{tripathi1975asymmetric}.
\begin{align}
Z_{c,a}&= \frac{\omega}{\beta_c}\frac{\left(\mathcal{L}_a+\mathcal{M}\right)\left(\mathcal{L}_b+\mathcal{M}\right)-\mathcal{M}^2}{\mathcal{L}_b+\mathcal{M}+\mathcal{M}R_c}
\\
Z_{c,b}&= \frac{R_c \omega}{\beta_c}\frac{\left(\mathcal{L}_a+\mathcal{M}\right)\left(\mathcal{L}_b+\mathcal{M}\right)-\mathcal{M}^2}{\left(\mathcal{L}_a+\mathcal{M}\right)R_c+\mathcal{M}}
\\
Z_{\pi,a}&= \frac{\omega}{\beta_\pi}\frac{\left(\mathcal{L}_a+\mathcal{M}\right)\left(\mathcal{L}_b+\mathcal{M}\right)-\mathcal{M}^2}{\mathcal{L}_b+\mathcal{M}+\mathcal{M}R_\pi}
\\
Z_{\pi,b}&= \frac{R_\pi \omega}{\beta_c}\frac{\left(\mathcal{L}_a+\mathcal{M}\right)\left(\mathcal{L}_b+\mathcal{M}\right)-\mathcal{M}^2}{\left(\mathcal{L}_a+\mathcal{M}\right)R_\pi+\mathcal{M}}.
\end{align}
Finally, the port voltages can be evaluated by applying the following boundary conditions:
\begin{align}
[V_{\mathrm{IN}}-V_a(z=-l)]/Z_{\mathrm{L}a}&=I_a(z=-l)    \\
-V_b(z=-l)/Z_{\mathrm{L}b}&=I_b(z=-l)   \\ 
V_a(z=0)/Z_{\mathrm{L}a}&=I_a(z=0)   \\ 
V_b(z=0)]/Z_{\mathrm{L}b}&=I_a(z=0)    
\end{align}
where $V_{\mathrm{IN}}$ is the input voltage at port 1 and $Z_{\mathrm{L}b}$ and $Z_{\mathrm{L}a}$ are the load impedances. 

For a symmetric coupler, $\mathcal{L}_{a}=\mathcal{L}_{b} = \mathcal{L}$ and $\mathcal{C}_{a} = \mathcal{C}_{b}=\mathcal{C}$. Moreover, we assume $\mathcal{M}/\mathcal{L}\ll1$, and the propagation constants reduce to
\begin{align}
        \beta_c&=\omega\sqrt{\mathcal{L}\mathcal{C}}\\
        \begin{split}
            \beta_\pi&=\omega\sqrt{\mathcal{L}\mathcal{C}}\sqrt{1+\frac{2\mathcal{E}}{\mathcal{C}}+\frac{2\mathcal{M}}{\mathcal{L}}+\frac{4\mathcal{M}\mathcal{E}}{\mathcal{L}\mathcal{C}}}\\
            & \approx \omega\sqrt{\mathcal{L}\mathcal{C}}\sqrt{1+2\mathcal{E}/\mathcal{C}}\end{split}
\end{align}
and similarly, the impedances for the $c$ and $\pi$ modes
\begin{align}
    Z_\mathrm{c}&=\sqrt{\frac{\mathcal{L}}{\mathcal{C}}} \\
    \begin{split}
    Z_\pi&= \sqrt{\frac{\mathcal{L}}{\mathcal{C}}}\frac{\sqrt{1+\frac{2\mathcal{E}}{\mathcal{C}}+\frac{2\mathcal{M}}{\mathcal{L}}+\frac{4\mathcal{M}\mathcal{E}}{\mathcal{L}\mathcal{C}}}}{1+2\mathcal{E}/\mathcal{C}}\\
&\approx \sqrt{\frac{\mathcal{L}}{\mathcal{C}}}\frac{1}{\sqrt{1+2\mathcal{E}/\mathcal{C}}}.
\end{split}
\end{align}
A signal injected through the input port is a superposition of the two modes and the energy propagates through the coupled structures, shuttling between the two lines with a periodicity $l_\mathrm{\pi}=\pi/\Delta\beta$. Here
\begin{equation}\label{deltabeta}
    \Delta\beta = \beta_\pi-\beta_c \approx \omega \sqrt{\mathcal{LC}} \left( \sqrt{1+\mathcal{E /C}} -1 \right).
\end{equation}  
From Eq. \ref{deltabeta}, the minimum length required for 50:50 forward coupling is
\begin{equation} \label{min-length}
    l_{\pi/2,\mathrm{sc}} = \frac{\pi}{2} \frac{1}{\Delta \beta}\approx \frac{\lambda_c}{4} \frac{1}{\sqrt{1+\mathcal{E /C}} -1},
\end{equation}  
where $\lambda_c$ is the guided wavelength for the common mode. Note that in the high-inductance regime, the coupling is mostly determined by the kinetic inductance and the capacitance terms. 

\begin{figure*}
    \centering
    \includegraphics{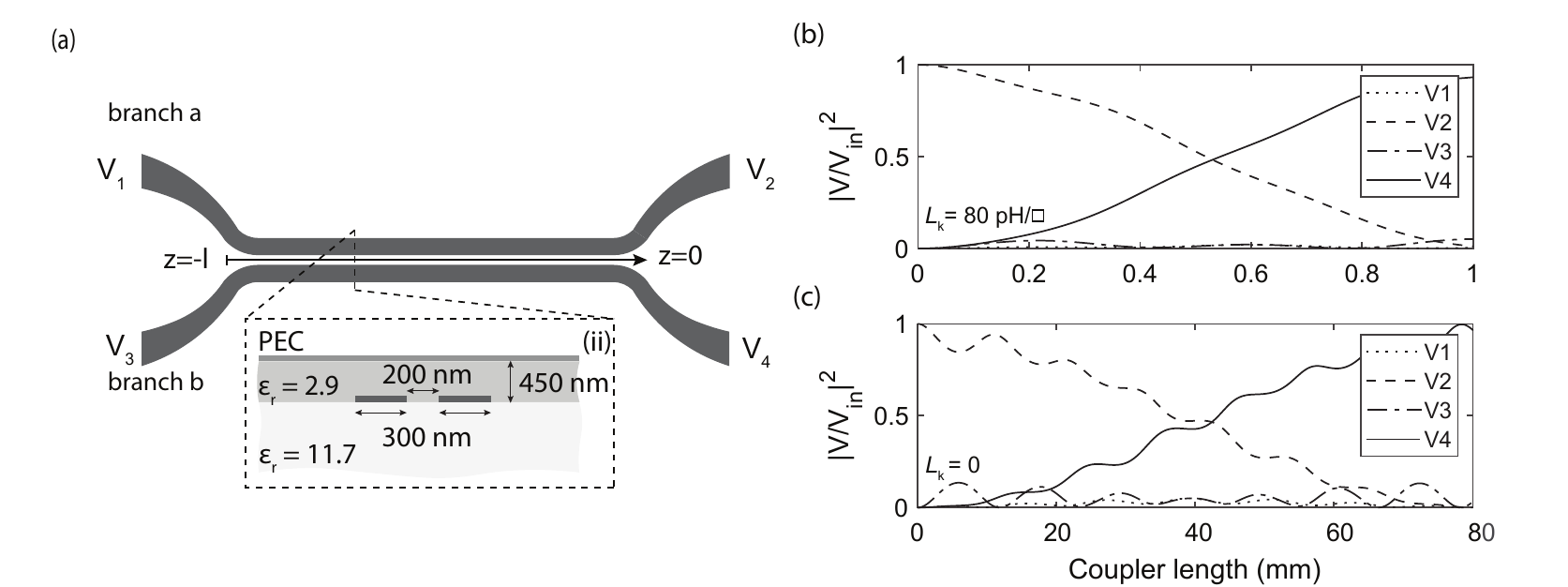}
    \caption{Analytical model of the coupling section. (a) Schematic of the coupling section. The ground layer was treated in simulation as a perfect electric conductor (PEC) (b) Voltage output from the coupling section for superconducting microstrip with $L_{\mathrm{k}}=80\,\mathrm{pH}$ per square. (c) Voltage output from the coupling section for normal conductor ($L_{\mathrm{k}}=0$).}
    \label{coupling_section_vs_Lk}
\end{figure*}

In Fig. \ref{coupling_section_vs_Lk} we show the calculation for a symmetric coupling section (Fig. \ref{coupling_section_vs_Lk}(a)). A $5\,\mathrm{GHz}$ input signal ($V_{\mathrm{IN}}$) injected into Port 1 of a side-coupled stripline with thin-film superconductors ($L_{\mathrm{k}}=80\,\mathrm{pH}$ per square) takes $539\,\mathrm{\mu m}$ to couple $50\%$ of the power to the coupled branch (Fig. \ref{coupling_section_vs_Lk}(b)). A normal conductor side-coupled stripline architecture can achieve 50:50 forward coupling (Fig. \ref{coupling_section_vs_Lk}(c)), but $\approx 40\,\mathrm{mm}$ coupling length is required. This represents an almost two-orders-of-magnitude size reduction.

\section{Microwave Simulation and Design}
\subsection{Generalities}
To perform microwave response simulations we used Sonnet EM from SonnetLab \footnote{Certain commercial equipment, instruments, or materials are identified in this paper to facilitate understanding. Such identification does not imply recommendation or endorsement by NIST, nor does it imply that the materials or equipment that are identified are necessarily the best available for the purpose. }. Note that, for simplicity, unless otherwise specified all the normal metal layers have been simulated as perfect electric conductors (PEC).  In actual experiments, the metallic ground does contribute non-negligible losses. This, however, can be minimized by using superconducting ground planes, such as Al at sub-1K temperature. In Table \ref{table:sim_param} we report the general material parameters used in the simulations.
\begin{table}[h]
    \centering
    \begin{tabular}{c|c}
       Substrate thick.  & $500\,\mathrm{\mu m}$   \\
       Substrate $\epsilon_{r}$  & 11.7 \\
       Dielectric thick $\epsilon_{r}$ & $450\,\mathrm{nm}$ \\
       Dielectric $\epsilon_{r}$ & 2.9 \\
       Au conductivity & $4.09 \times 10^7 \,\mathrm{S/m}$ \\
       Au thick. &$60\,\mathrm{nm}$ \\
    \end{tabular}
    \caption{Material parameters used in the simulations.}
    \label{table:sim_param}
\end{table}

\subsection{Single nanowire microstrip}

\begin{figure}[h]
    \centering
    \includegraphics[width=\linewidth]{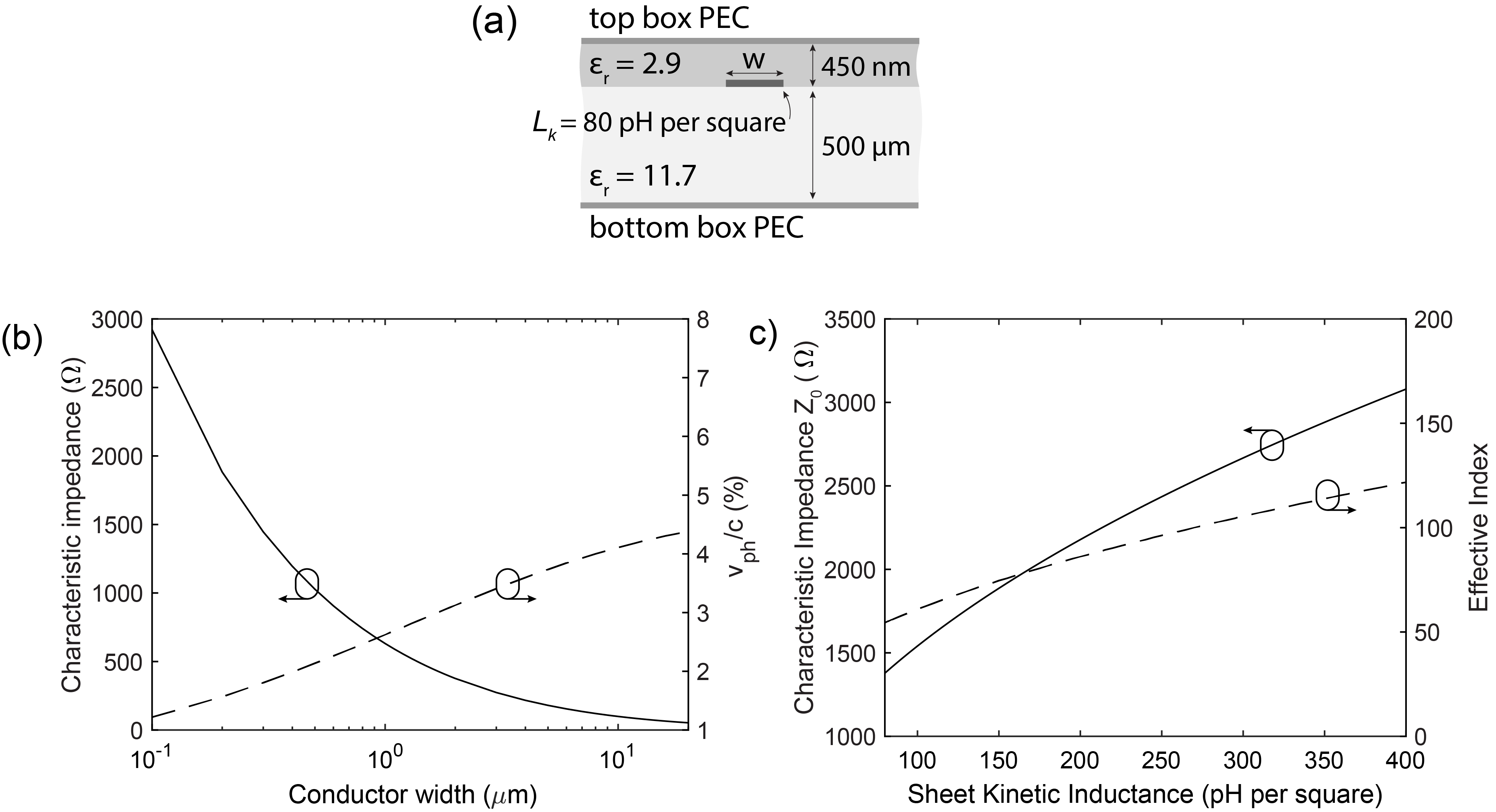}
    \caption{Single nanowire microstrip simulation setup and result. (a) Stack and geometry setup as in Sonnet EM. (b) Characteristic impedance and phase velocity fraction versus conductor width for a sheet kinetic inductance $L_{k}=80\,\mathrm{pH}$ per square. (c) Characteristic impedance and effective index for a $320\,\mathrm{nm}$ wide microstrip line, as a function of the kinetic inductance.}
    \label{s_fig1}
\end{figure}

The material stack and the geometry were set as in Fig.  \ref{s_fig1}(a). The results for the characteristic impedance $Z_0$ and phase velocity fraction versus the width of the conductor for a sheet kinetic inductance $L_{k}=80\,\mathrm{pH}$ per square are shown in Fig.  \ref{s_fig1}(b). Fig.  \ref{s_fig1}(c) shows the characteristic impedance and effective index for a $320\,\mathrm{nm}$ wide microstrip line, for several values of the kinetic inductance.

\subsection{Coupled nanowire microstrip lines} \label{even_odd_imp}

\begin{figure*}
    \centering
    \includegraphics[width=\linewidth]{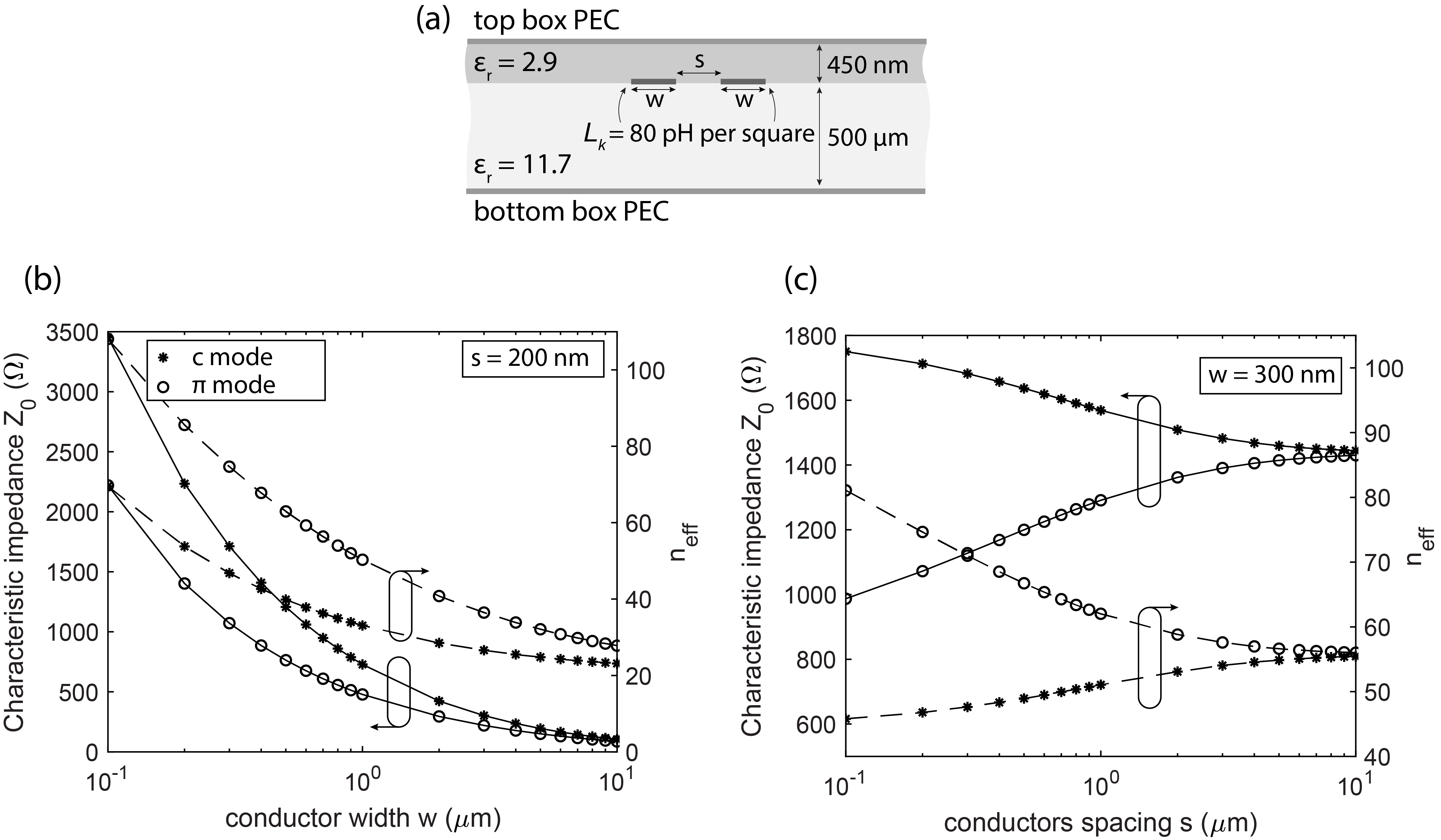}
    \caption{Characteristic impedance and effective indices for coupled microstrip lines.  (a) Stack and geometry setup as in Sonnet EM. (b) Characteristic impedance and effective index for the common and differential mode versus conductor width (with constant gap) or gap (with constant width), for a sheet kinetic inductance $L_{k}=80\,\mathrm{pH}$ per square. }
    \label{s_even_odd}
\end{figure*}

The material stack and the geometry were set as in Fig.  \ref{s_even_odd}(a). The results for the characteristic impedances $Z_0$ and effective indices are shown in Fig.  \ref{s_even_odd}(b). We performed the simulation for several conductor widths $w$ (keeping the gap constant) and gap spacings $s$ (keeping the width constant), starting from the design values, $w=300\,\mathrm{nm}$ and $s=200\,\mathrm{nm}$, which were selected to minimize challenges in the fabrication. The sheet kinetic inductance was $L_{k}=80\,\mathrm{pH}$ per square.

\subsection{Simulation of the full device} 
The device was simulated as a netlist starting from full simulations of single components: the 4 port coupler, the 2 port taper, and the coplanar waveguide feedline. This simulation does not account for specific element-to-element transitions. For example, the transmission line arcs connecting the 4 port coupler with each taper or the transition between the microstrip architecture to the CPW feed lines are not captured by the simulated layout. As the isolation and directivity of a parallel line coupler depends on the real geometric layout of the coupler \citep{ikalainen1987wide}, the simulated value may be inaccurate. This might also explain the significant discrepancy with the measured isolation.

\subsection{Design of the impedance-matching taper}

\begin{figure}[h]
    \centering
    \includegraphics{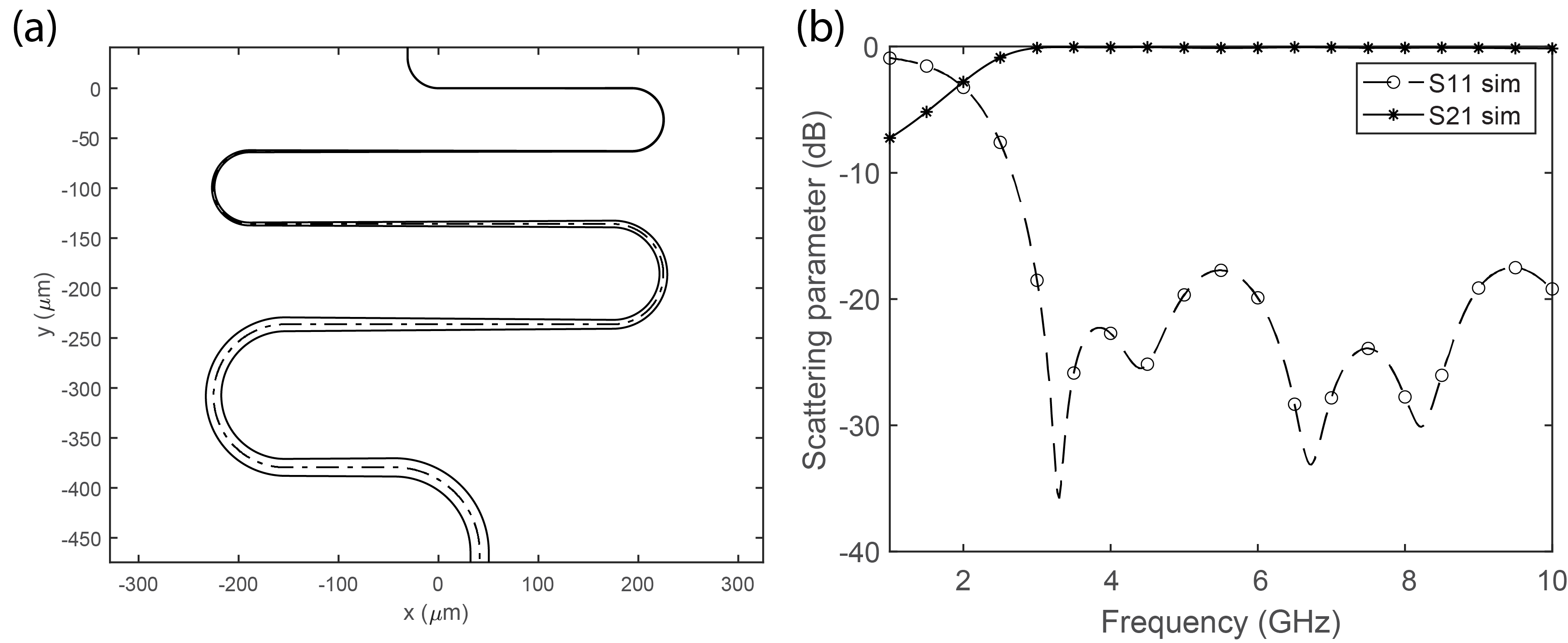}
    \caption{Taper: layout and microwave dynamics simulation. (a) Layout of the designed taper (b) S parameters simulation of the taper assuming $L_{k}=80\,\mathrm{pH}$ per square.}
    \label{s_taper}
\end{figure}

We designed a Klopfenstein impedance matching taper following the works in Ref. \citep{zhu2019superconducting,klopfenstein1956transmission,zhu2019resolving}. The taper transforms the impedance of a transmission line, minimizing reflections. One end of the taper had a width of $300\,\mathrm{nm}$ to match the nanowire transmission line at the coupler section. The other end  of the taper is designed to provide a $50\,\mathrm{\Omega}$ characteristic impedance to match the room-temperature  testing electronics. For the microwave environment of this transmission line, $50\,\mathrm{\Omega}$ is obtained with a width of $\approx 15\,\mathrm{\mu m}$. The taper had a total length of $\approx 1.97\,\mathrm{mm}$ in 5214 sections, with $ \approx 1178$ squares, for a total inductance of $\approx 94\,\mathrm{nH}$ assuming $L_{k}=80\,\mathrm{pH}$ per square. The cut-off frequency was designed to be $f_{\mathrm{co}}=2.5\,\mathrm{GHz}$. In Fig. \ref{s_taper}(a), we show the layout of the taper. In Fig. \ref{s_taper}(b), we show the simulation of the taper. The $-3\,\mathrm{dB}$ cutoff frequency is at $2.5\,\mathrm{GHz}$.

\subsection{Parallel line coupler with asymmetric dielectrics}
\begin{figure}
    \centering
    \includegraphics{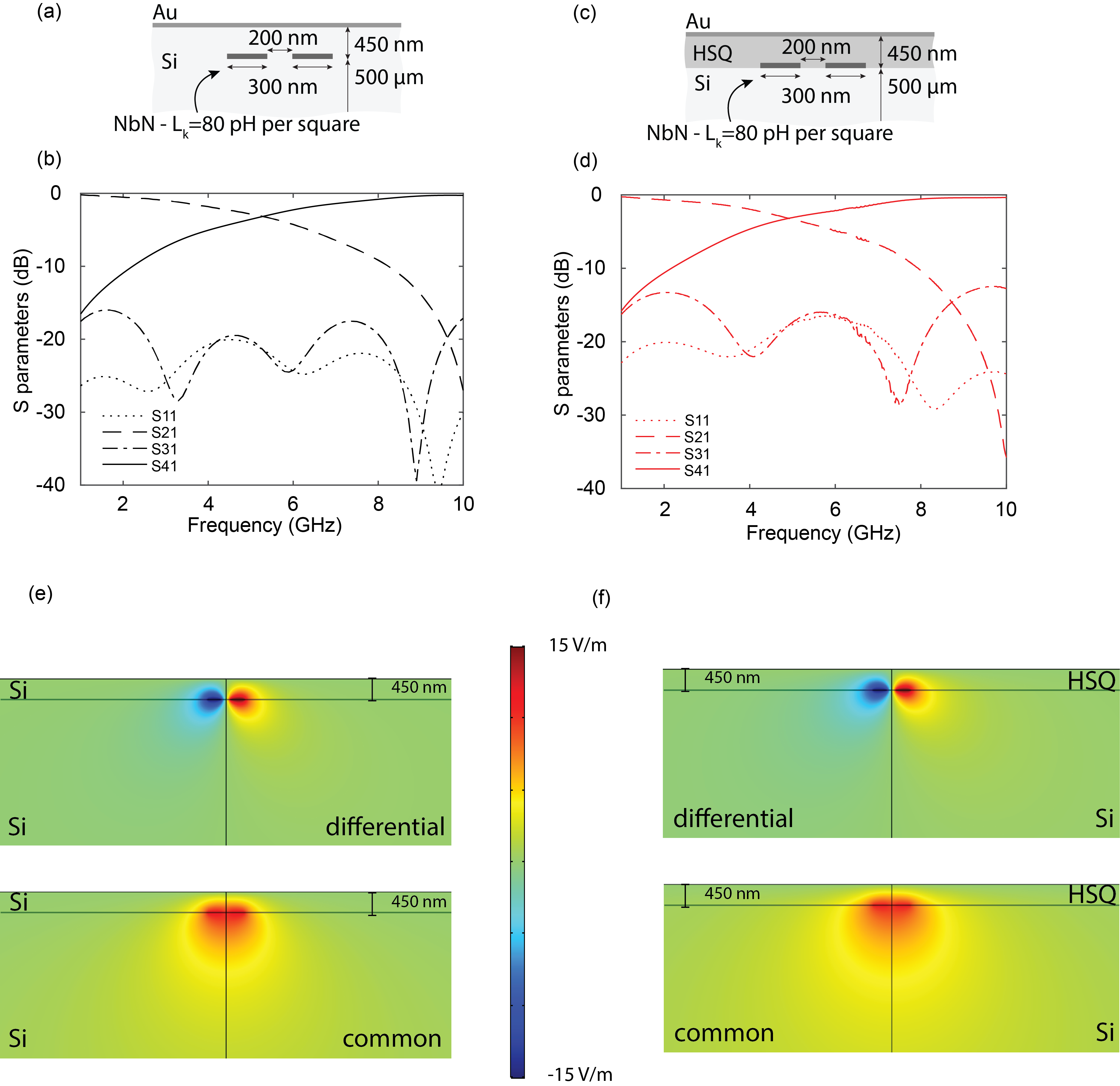}
    \caption{Cross section and S-parameters simulation of side-coupled coupler with symmetric dielectric (a)-(b), and asymmetric dielectric (c)-(d). Apart from a slight shift in the frequency response, dictated by the different impedance/index environment, there are no significant differences. (e) and (f) show the simulation of the common and differential modes for the two cases. }
    \label{homo}
\end{figure}
In printed circuit boards, the performance of topside coupled-microstrip couplers is degraded due to asymmetry of the dielectrics (e.g. air/FR4). In this section we assess the impact of asymmetric dielectrics on our integrated coupler. 
We simulate the same coupler architecture described in the main text (Fig. \ref{homo}(b)) with a symmetric dielectric, replacing the HSQ layer with intrinsic silicon. The characteristic-impedance of the architecture in Fig. \ref{homo}(a) resulted as $Z_0=965\,\mathrm{\Omega}$. We used this value to terminate each port of the homogeneous-dielectric coupler. In Fig. \ref{homo}(c) and (d) we compare the S-parameters of the symmetric dielectric coupler (c) with the original asymmetric dielectric coupler (d). This analysis shows that  apart from frequency shifts, induced by the different microwave environments, there are no significant differences. This is confirmed by the common and differential mode profile simulated with finite element modeling (COMSOL). This qualitatively indicates the absence of effects due to dielectric discontinuities and asymmetry.

\subsection{Qualitative simulation of a broadside coupler architecture}
\begin{figure}
    \centering
    \includegraphics{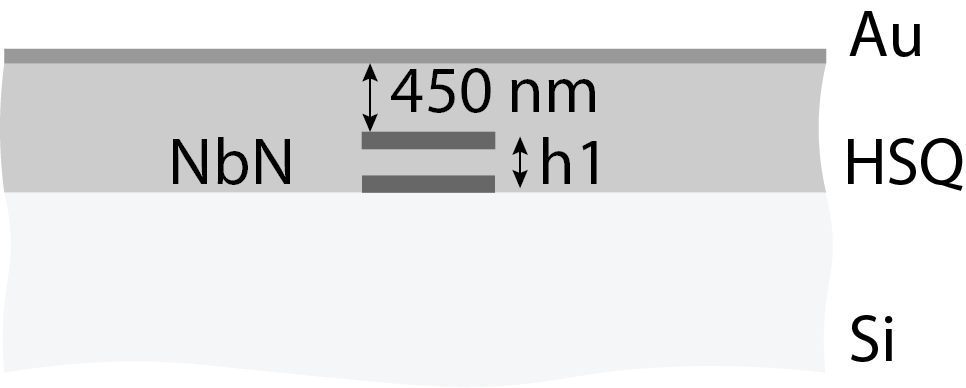}
    \caption{Broadside coupler architecture: cross-section sketch}
    \label{broadside}
\end{figure}

A broadside coupler architecture, where two nanowires are stacked vertically as shown in Fig. \ref{broadside} can be used to increase the coupling capacitance and obtain an even shorter coupling length. To explore the impact of realizing a broadside coupler using nanowire transmission lines we simulated the S-parameters of the architecture in Fig. \ref{broadside}, for several values of the dielectric thickness $h_1$. Other parameters were left unchanged to have minimal variation compared to the side-coupled nanowire architecture explored in this work. In Table \ref{table_broad} we report the value of the 50:50 coupling frequency. While decreasing $h_1$, the 50:50 coupling frequency also decreases. As expected, the increase of coupling capacitance reduces the minimum coupling length (compared to the side-coupled coupler).

\begin{table}[h]
    \centering
    \begin{tabular}{c|c}
       $h_1$  & $f_{50:50}$ \\ \hline
      $50\,\mathrm{nm}$ \quad  & \quad $2.5\,\mathrm{GHz}$ \\
      $100\,\mathrm{nm}$ \quad & \quad $3.6\,\mathrm{GHz}$ \\
      $180\,\mathrm{nm}$ \quad & \quad $5.2\,\mathrm{GHz}$ \\
    \end{tabular}
    \caption{Broadside coupler design 50:50 coupling frequency.}
    \label{table_broad}
\end{table}

\section{Measurement setups, calibration and characterization}

\subsection{Estimation of $I_{sw}$ and $T_c$}
\begin{figure}[h]
    \centering
    \includegraphics{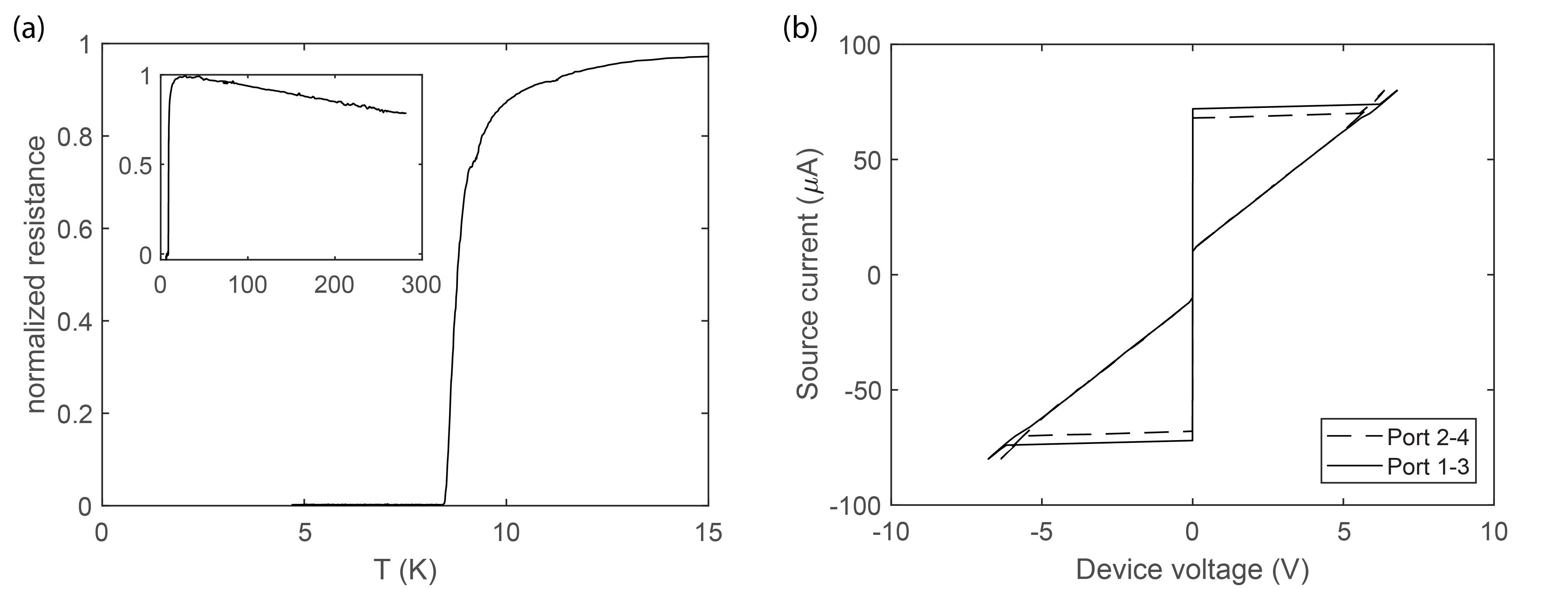}
    \caption{(a) Characterization of the critical temperature of a bare film from the same wafer used to fabricate the coupler. (b) Characterization of the switching currents of the coupler branches.}
    \label{dc}
\end{figure}

In Fig. \ref{dc}(a) we show the (normalized) $R$ vs. $T$, which we used to characterize the critical temperature $T_c=8.78\,\mathrm{K}$, taken at $50\%$ of the superconducting transition, and the residual resistance ratio $RRR=0.8$. The characterization was performed in liquid He on a blanket film from the same wafer used to fabricate the device. The sheet resistance was measured at room temperature with a four-point probe setup as $R_s=360\,\mathrm{\Omega}$ per square.

After fabrication we measured the switching current of the coupler device at $T=1.3\,\mathrm{K}$. For the input/transmission branch (Port 1-2) we measured $I_{sw}=72\,\mathrm{\mu A}$. For the isolation/coupling branch (Port 3-4), $I_{sw}=68\,\mathrm{\mu A}$. The variation in switching current is $4\,\mathrm{\mu A}$ and reflects a slight electrical asymmetry that may be due to fabrication or local imperfections of the film.

\subsection{Measurement setups for microwave characterization}
\begin{figure}[h]
    \centering
    \includegraphics{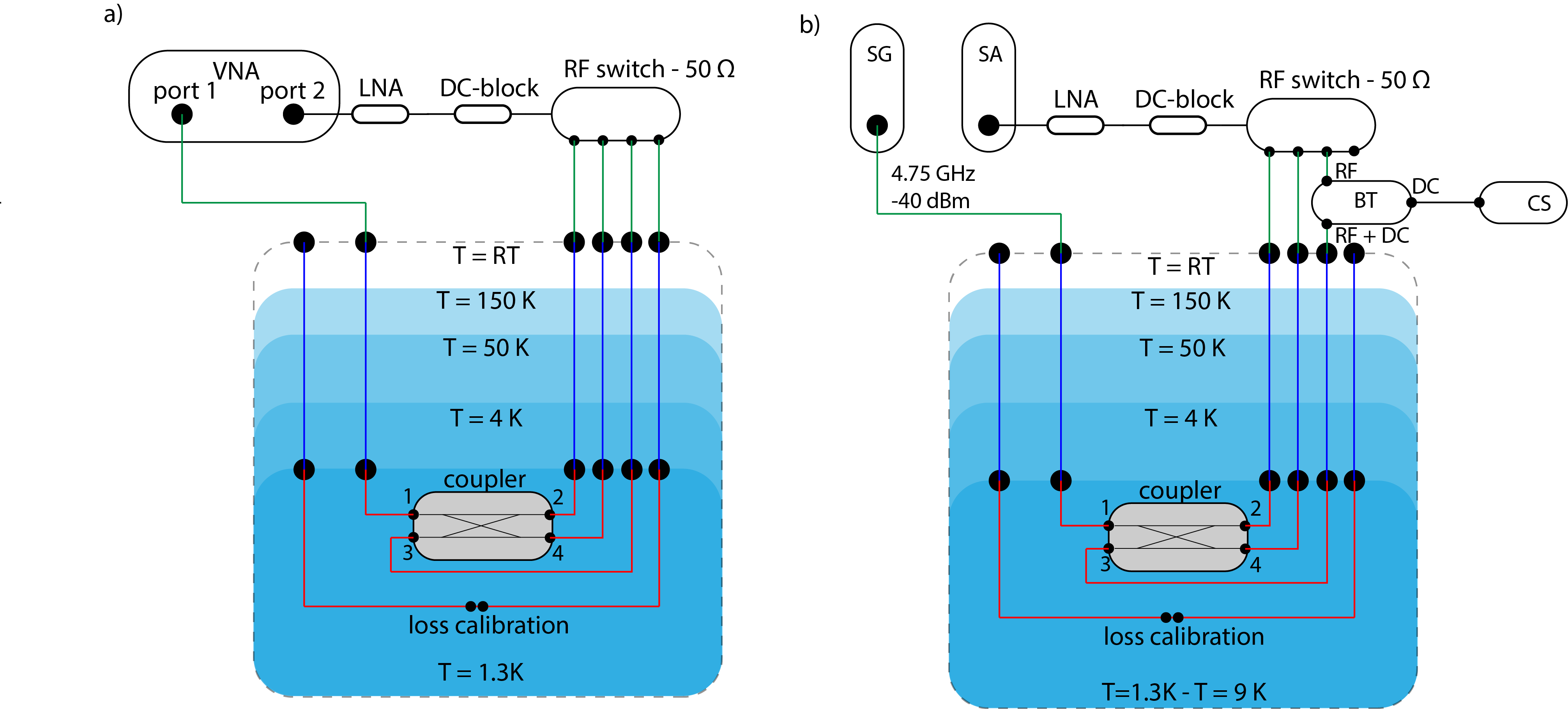}
    \caption{Measurements setups. (a) Measurement setup used to characterize the microwave response of the coupler and to perform the loss calibration. (b) Measurement setup used to characterize the coupling-point tunability.}
    \label{setup}
\end{figure}

In this section we discuss the measurement setups used for the microwave characterization.
Fig.  \ref{setup}(a) shows a sketch of the measurement setup used for the characterization of the S parameters and to perform the cable and connector loss calibration. The packaged device was cooled down to $1.3\,\mathrm{K}$ in a closed-cycle cryostat (ICE Oxford) \footnote{Certain commercial equipment, instruments, or materials are identified in this paper to facilitate understanding. Such identification does not imply recommendation or endorsement by NIST, nor does it imply that the materials or equipment that are identified are necessarily the best available for the purpose. }. The input of the coupler was connected to the Port 1 of the vector network analyzer (VNA) (Keysight N5242A) \footnotemark[2]. The other three ports of the coupler (isolation, transmission and coupling) were connected to an RF switch at room temperature (MSP4TA-18-12+) \footnotemark[2]. The common port of the switch was connected to the Port 2 of the VNA, through a DC block and a room-temperature low-noise amplifier (RF Bay LNA-8G) \footnotemark[2]. The cables illustrated with same colors are of the same length and from the same manufacturer. The loss calibration was performed at every cooldown and consisted of a transmission measurement (S21) of the input and output cable assembly, connected together; the device was not included. To connect the cables, we used a barrel female-to-female straight SMA connector, anchored to the $1.3\,\mathrm{K}$ stage. Therefore, the calibration does not account for losses and reflections from the PCB, RF box, and wire bonds. 

Fig. \ref{setup}(b) shows the measurement setup used for the characterization of the S-parameter tunability. At the input, the VNA was replaced by a signal generator (Windfreak SynthHD PRO) \footnotemark[2] and, at the output, by a spectrum analyzer (Aglient N9030A)\footnotemark[2]. The tuning DC current was supplied with a current source to the isolation port of the coupler through a bias tee (ZFBT-6GW+) \footnotemark[2]. The temperature of the $1.3\,\mathrm{K}$ stage was controlled by a heater. When measuring the coupling-point tunability with temperature, we allowed for a 5 minute stabilization time to avoid temperature fluctuation during acquisition.

\subsection{Calibration and raw traces}
In this section we show the calibration data and the measured S parameters before correction. The calibration included the cables at room temperature, the cables from room temperature to the devices at $1.3\,\mathrm{K}$, the intermediary connectors, the RF switch, the DC block, and the LNA. 
\begin{figure}[h]
    \centering
    \includegraphics{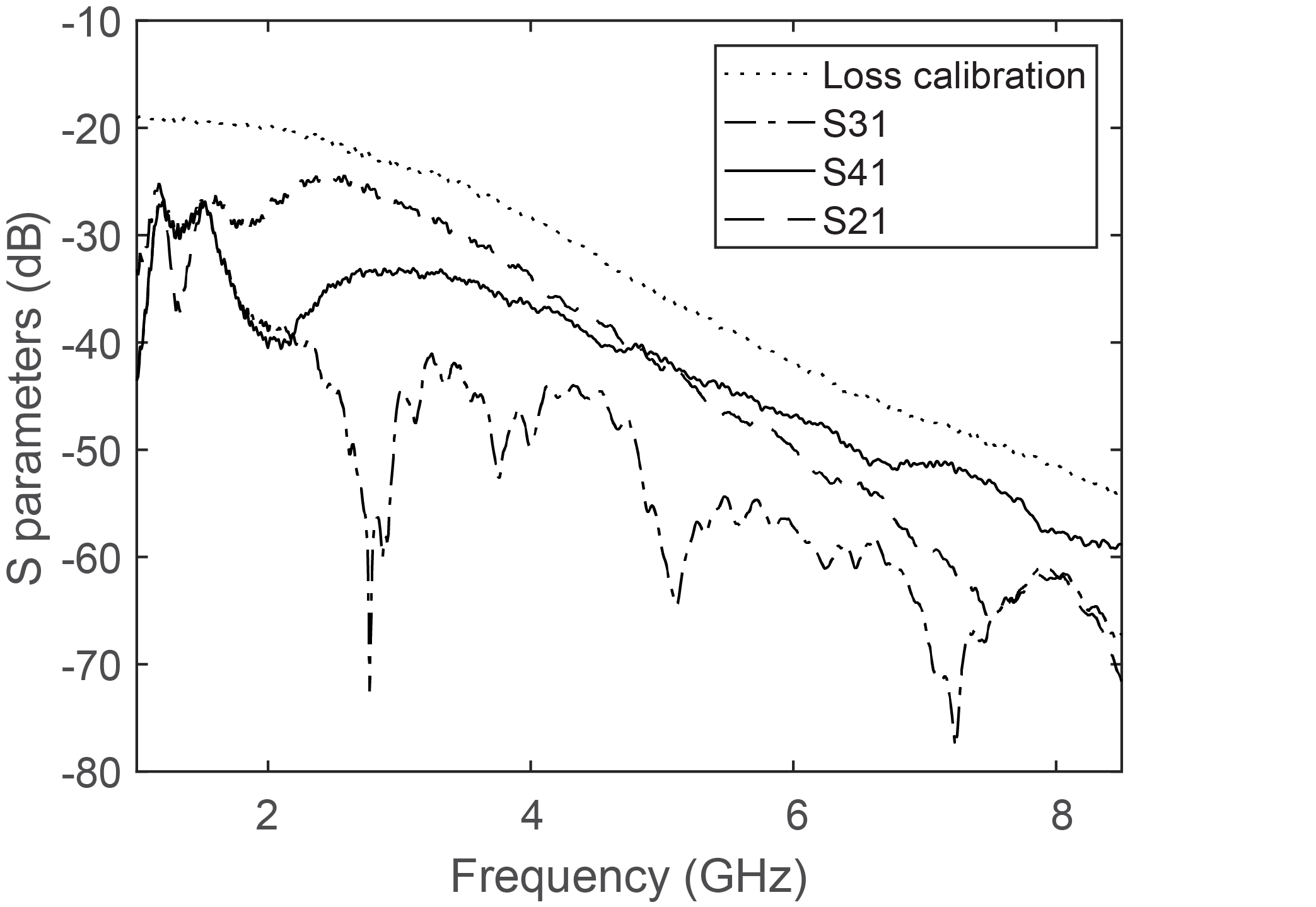}
    \caption{Loss calibration data and coupler S parameters before correction.}
    \label{calibration}
\end{figure}

\subsection{Coupler phase}
We measure the phase of the signals out of the coupler. The phase difference is fixed at $90^\circ$ due to the symmetry of the coupler \citep{ikalainen1987wide}. This is confirmed by the the analytic model and the simulation (in the taper passband). 

\begin{figure}
    \centering
    \includegraphics{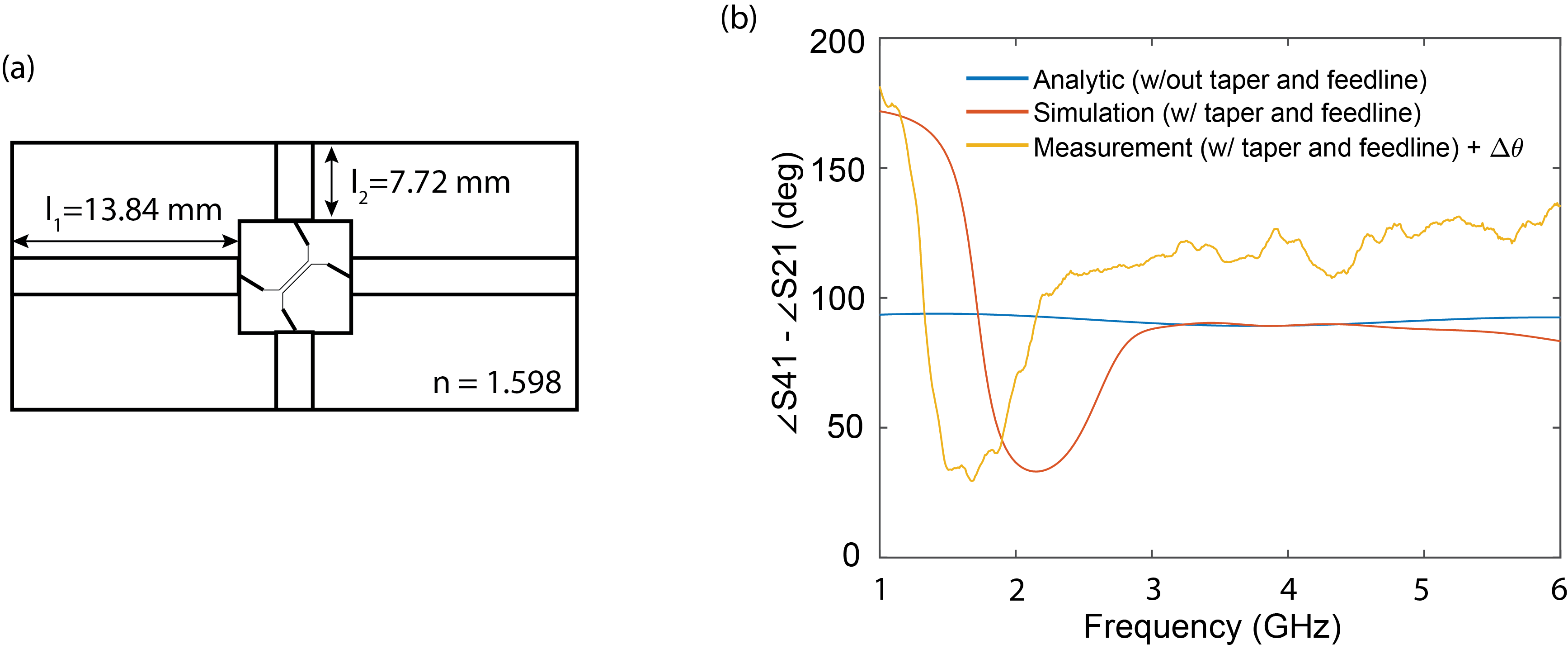}
    \caption{(a) Sketch of the PCB used for measurements. (b) Phase shift of the signals at the two output ports.}
    \label{pcb}
\end{figure}

The PCB used for the measurement has asymmetric trace lengths (Fig. \ref{pcb}(a)). In order to calculate the phase of the signals out of the coupler we need to take into account the phase shift due to the PCB trace lengths difference. The phase shift $\Delta \theta (f)$ is estimated as follows

\begin{equation}
    \Delta \theta (f) = 360^\circ f \frac{l_1-l_2}{v_{\mathrm{ph}}} 
\end{equation}
where $v_{\mathrm{ph}}=c/n$ is the phase velocity, $n$ is the effective index of the PCB transmission lines. 
When taking into account the PCB phase shift, the phase difference is found to be higher than $90^\circ$. The shape of the measured phase resembles the one obtained with the simulation. Therefore, we attribute this discrepancy to additional unaccounted phase shifts.

\subsection{Taper characterization}
\begin{figure}
    \centering
    \includegraphics{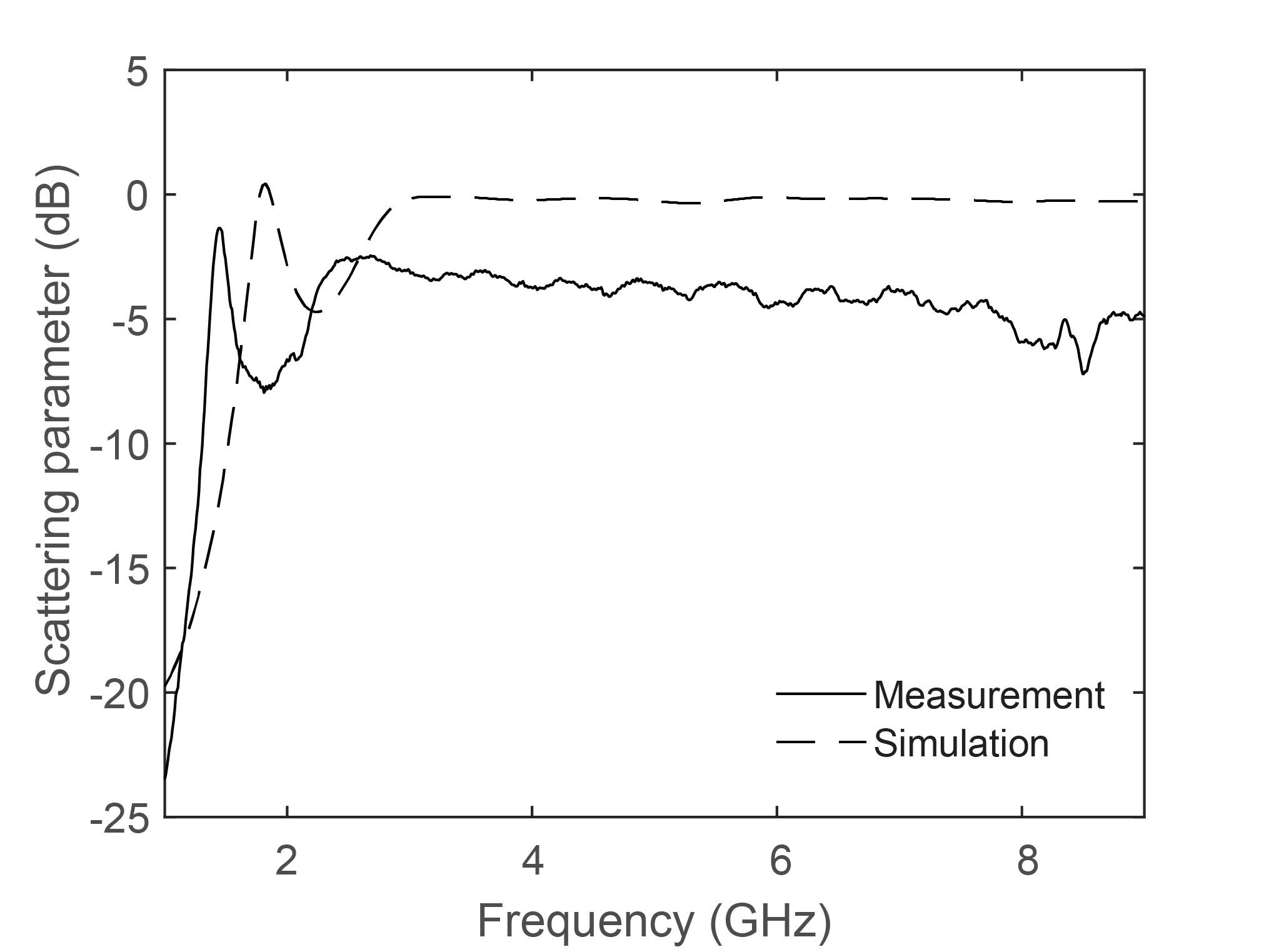}
    \caption{Characterization of the Taper-Line response and comparison with the corresponding simulation.}
    \label{taper_meas}
\end{figure}

In this section we show the characterization of the impedance-matching taper. In the main text we argued that the taper, which is indispensable to perform the measurement with $50\,\Omega$ electronics (as in this case), was deviating from the design and could be a source of added reflection that could degrade the measured performances.
In a separate cooldown, we measured a two-port device consisting of only one of the branches of the coupler. This device was fabricated in the same run as the coupler and included two CPW microwave launchers, two impedance-matching tapers and one $520\,\mathrm{\mu m}$ long nanowire. We will refer to this device as \textit{Taper-Line}, for convenience. In Fig.  \ref{taper_meas} we show the transmission S21 of the Taper-Line and a corresponding simulation. Overall, compared to the design, the measured microwave response was shifted to lower frequency, with a cutoff frequency  $f_{co}\approx 2\,\mathrm{GHz}$. The magnitude difference and the presence of in-band frequency dependence (the transmission decreases for higher frequencies) are attributed to device level loss induced by the presence of normal conductors (Au top ground and CPW lines).  

\section{Fabrication process} \label{AppD}
A $\approx 7\,\mathrm{nm}$-thick NbN film was sputter deposited \citep{dane2017bias} on a $2\,\mathrm{cm}$ x $2\,\mathrm{cm}$ high-resistivity Si substrate. The room temperature sheet resistance was $R_s=360\,\mathrm{\Omega}$ per square, the residual resistance ratio $RRR=0.8$ and the critical temperature $T_C = 8.8\,\mathrm{K}$. 

The sheet kinetic inductance at $T=0\,\mathrm{K}$ can be estimated as \citep{tinkham2004introduction} $ L_{k}= \left(\hbar R_s\right)/\left( RRR \pi \Delta_0\right)\approx 70.5\,\mathrm{pH}$ per square, where $\Delta_0$ is zero-temperature superconducting gap. Film storage and fabrication process may induce the degradation of the superconducting properties of the film, resulting in a higher sheet resistance and a lower critical temperature. For a device fabricated with a similar NbN film, we recently reported $L_{k} \approx 80 \,\mathrm{pH}$ per square \citep{zhu2019superconducting}, approximately $10\%$ higher than the theoretical estimation. 

\begin{figure}
    \centering
    \includegraphics{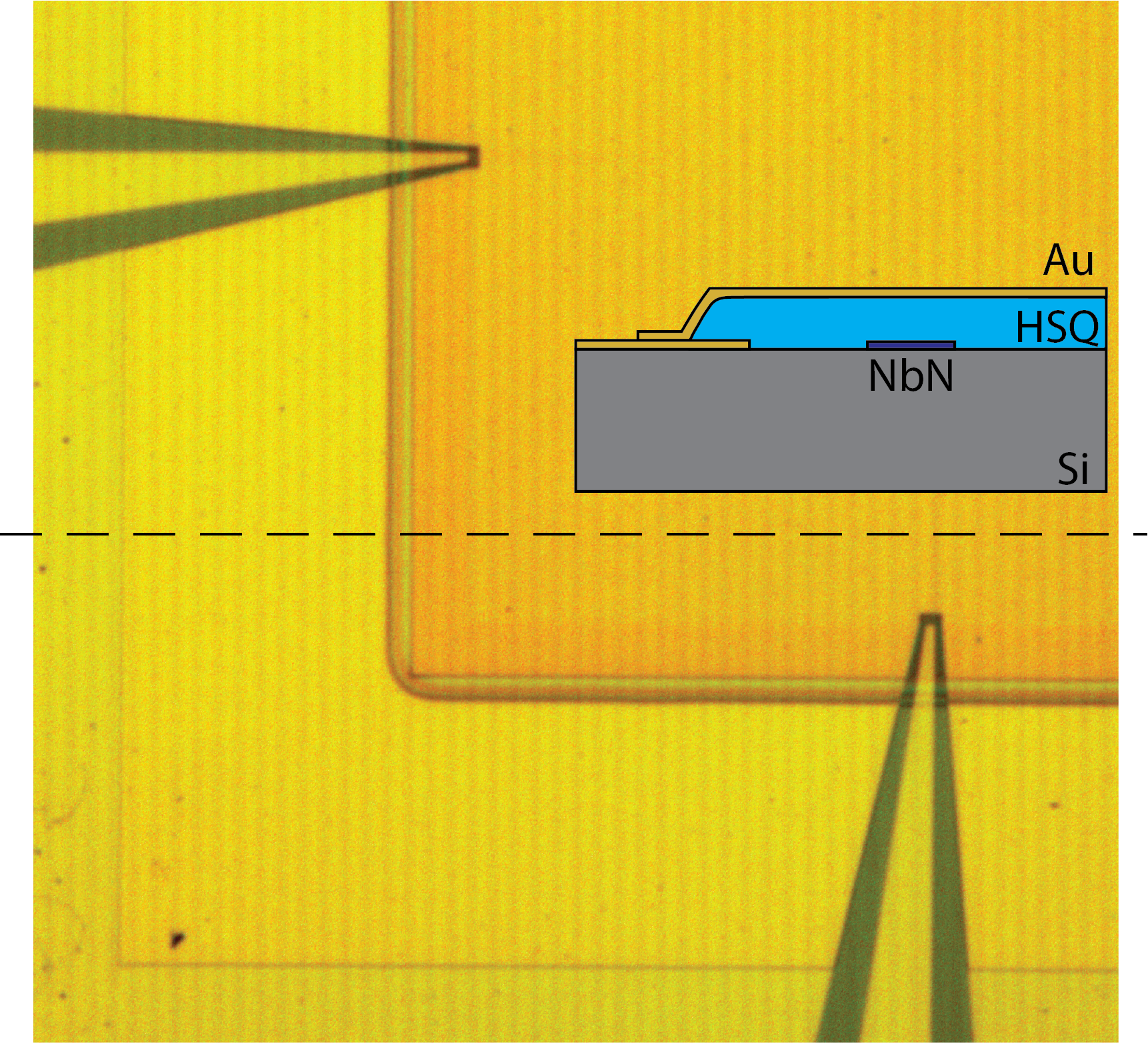}
    \caption{Topside ground to feedline ground connection with cross section sketch.}
    \label{gndtopside}
\end{figure}

We started with the fabrication of $50\,\mathrm{\Omega}$ coplanar waveguide (CPW) feed lines using direct writing photolithography (DWL), followed by gold evaporation and liftoff. We patterned the nanowire transmission lines and Klopfenstein tapers by aligned negative-tone electron beam lithography, using ma-N 2401 \citep{toomey2019investigation}, and we transferred the patterns into the NbN through reactive ion etching with CF$_4$. We completed the microstrip structure, by patterning a $450\,\mathrm{nm}$ hydrogen silsequioxane (FOx-16) dielectric spacer, having $\epsilon_r=2.9$, using a purposely designed low-contrast electron beam lithography process \citep{zhu2018scalable}. Lastly, we fabricated the top ground, with aligned DWL, followed by gold evaporation and liftoff. The contact between the topside ground and the feedline ground is achieved by exploting the sloped sidewalls of the dielectric brick resulting from the low contrast process (Fig. \ref{gndtopside}).
After fabrication, the width of the lines was $320\,\mathrm{nm}$ while physical separation was reduced to $180\,\mathrm{nm}$, due to proximity effect. The superconducting transition of the fabricated device was observed at $T_c\sim 8\,\mathrm{K}$, reflecting film degradation during fabrication. Each chip contained $4$ coupler dies and several other test structures. The measurements reported in this paper are all obtained from one device. The micrographs are from a test structure fabricated with exactly the same process on the same starting wafer. A die designed for $10\,\mathrm{GHz}$ was also fabricated and tested, with worse performance due to limitations of the testing setup at higher frequencies. The packaging consisted of a custom made gold-plated copper box with a copper-plated Rogers 4003 PCB.
\section{Current in the nanowire} \label{AppCurrent}

To characterize the microwave response of the coupler, we used a nominal signal power of $-40\,\mathrm{dBm}$ from the room temperature electronics. We estimate that the power at the device’s input port, after accounting for system attenuation, was less than $-60\,\mathrm{dBm}$. Assuming an ideal broadband impedance transformer, with perfect matching and transmission, we can assume the following relation between the voltage applied at the low impedance end, $V_{\mathrm{L}}$, and the current at the high impedance end $I_{\mathrm{H}}$ \citep{zhu2019superconducting}
\begin{equation}
I_{\mathrm{H}}=\frac{V_{\mathrm{L}}}{\sqrt{Z_{\mathrm{L}}Z_{\mathrm{H}}}}.
\end{equation}
For $Z_{\mathrm{L}}=50\,\mathrm{\Omega}$ and $Z_{\mathrm{H}}=1446\,\mathrm{\Omega}$, a $-60\,\mathrm{dBm}$ signal converts to a $V_{\mathrm{pk,Z_{\mathrm{L}}}}=316.2\,\mathrm{\mu V}$ and $I_{\mathrm{pk,Z_{\mathrm{H}}}}=1.2\,\mathrm{\mu A}$. $I_{\mathrm{pk,Z_{\mathrm{H}}}}$ is much lower than the switching current of the wire. Therefore, we assume the influence of kinetic inductance non-linearities to be negligible for these measurements. 

\section{Kinetic inductance non-linearities}\label{AppGap}
The kinetic inductance strongly depends on carrier density, which can be tuned with a dc current and with temperature: $L_k=L_k(x=I_\mathrm{b}/I_\mathrm{d},t=T/T_\mathrm{c})$, where $I_\mathrm{d}$ is the depairing current. Note that in the main text we used $x'=I_\mathrm{b}/I_\mathrm{sw}$, where $I_\mathrm{sw}$ is the switching current.
The dependence of the kinetic inductance on current, in the fast relaxation regime \citep{clem2012kinetic} for $t=0.2$, can be described to an accuracy of $1\%,$ by 
\begin{equation}    \label{lkin_current}
    \frac{L_k(x,0.2)}{L_k(0,0.2)}=\left(\frac{1}{1-x^{2.27}}\right)^\frac{1}{2.27}.
\end{equation}
At the base temperature $T=1.3\,\mathrm{K}$ the curve at $t=0.2$ should be reasonably accurate.

\begin{figure}
    \centering
    \includegraphics{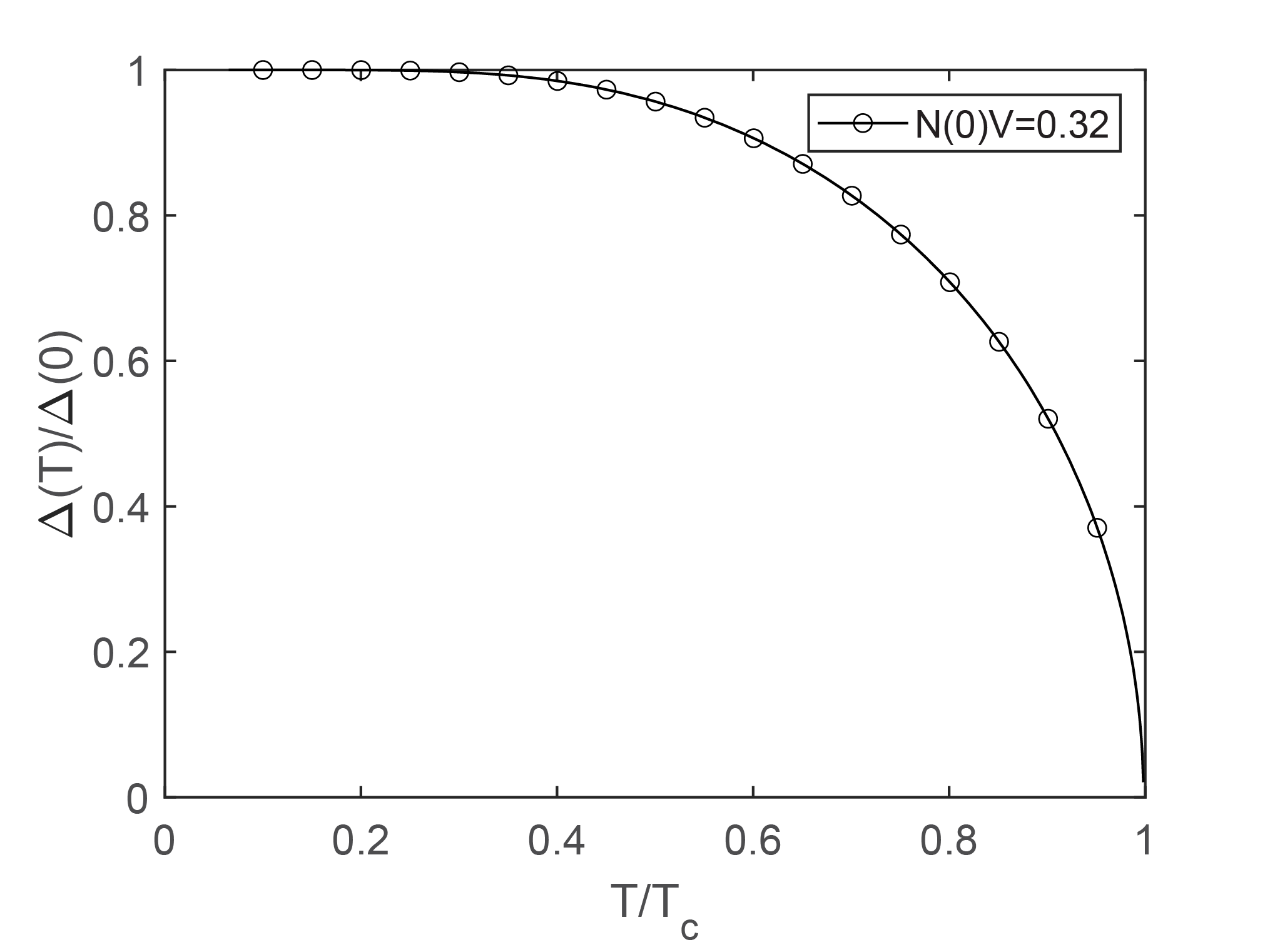}
    \caption{Result of the calculation of $\Delta(T)$ with $N(0)V=0.32$ and $T_{\mathrm{c}}=8\,\mathrm{K}$.}
    \label{deltat}
\end{figure}

The dependence of the kinetic inductance on temperature is described by \citep{santavicca2016microwave}
\begin{equation}
\frac{L_k(0,t)}{L_k(0,0)}=\frac{\Delta(0)}{\Delta(T) \tanh \left[ \Delta(T)/\left( 2k_{B}T\right) \right]}    
\label{lkin_t}.
\end{equation}
To simplify the calculation in the main text we made the assumption that $L_k(0,0.2) \approx L_k(0,0) $.
From BCS theory it is possible to derive an integral expression for the superconducting gap dependence on temperature \citep{tinkham2004introduction}
\begin{equation}
\frac{1}{N(0)V}=\int^{\hbar\omega_c}_{0} \frac{\tanh{\frac{1}{2}\beta\left(\xi^2 + \Delta^2\right)^{1/2}}}{\left(\xi^2 + \Delta^2\right)^{1/2}} d\xi   
\end{equation}
where $\beta^{-1}=k_BT$ ($k_B$ is the Boltzmann constant), $N(0)$ is the density of states at Fermi surface, $V$ is the BCS interaction potential, $\Delta$ is the superconducting gap, $\omega_c$ is the Debye frequency, and $\xi$ is the energy. This notation is inherited from the BCS pairing Hamiltonian formulation \citep{tinkham2004introduction}.
We first simplified the integral expression (by using the solution for the zero-temperature superconducting gap and for the critical temperature) to 
\begin{equation}
\frac{1}{N(0)V} \approx \int^{\frac{\Delta(0)}{\Delta(T)} \sinh{\frac{1}{N(0)V}}}_{0} \frac{\tanh{\frac{1.76}{2}\frac{\Delta(T)}{\Delta(0)}\left(1 + \zeta^2\right)^{1/2}}}{\left(1 + \zeta^2\right)^{1/2}} d\zeta
\end{equation}
where $\zeta=\xi / \Delta$. Then we numerically integrated and solved self-consistently by setting $N(0)V=0.32$ \citep{kang2011suppression,polakovic2018room}. This value holds for bulk NbN, so we might expect discrepancies when used for calculation with thin-film NbN, where a correction should be applied. The result of the calculation is shown in Fig.  \ref{deltat}.

\begin{figure}
    \centering
    \includegraphics{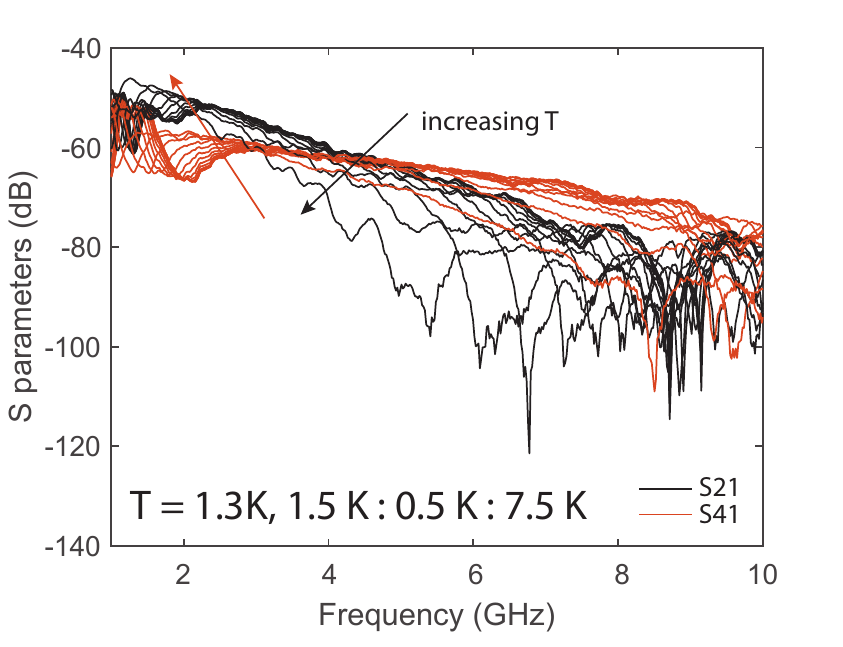}
    \caption{S parameters temperature tunability: raw data. }
    \label{rawtemp}
\end{figure}

The curves shown in Fig.  3 of the main text were obtained by modulating the kinetic inductance of the line according to Eq. \ref{lkin_current} and Eq. \ref{lkin_t}.

The S parameter tunability data reported in the main text were obtained using the setup of Fig. \ref{setup}(b). In Fig. \ref{rawtemp} we show the raw data for S21 and S41 with respect to the cryostat temperature, obtained through a VNA. This measurement shows the direct effect of the modulation temperature: the 50:50 coupling frequency shifts to lower frequencies while the parameters at the original 50:50 coupling frequency get modulated accordingly.

\bibliography{BiblioXY}